\documentclass[twocolumn]{aastex63}

\usepackage{bm}
\usepackage[T1]{fontenc}

\newcommand{\av}{a_\mathrm{V}}
\newcommand{\ac}{a_\mathrm{c}}
\newcommand{\ag}{a_\mathrm{g}}
\newcommand{\amon}{a_\mathrm{m}}
\newcommand{\xmon}{x_\mathrm{m}}
\newcommand{\mmon}{m_\mathrm{m}}
\newcommand{\df}{D_\mathrm{f}}
\newcommand{\kf}{k_\mathrm{f}}

\newcommand{\amax}{a_\mathrm{V}^\mathrm{max}}
\newcommand{\amin}{a_\mathrm{V}^\mathrm{min}}
\newcommand{\acmax}{a_\mathrm{c}^\mathrm{max}}

\newcommand{\org}{\texttt{org}}
\newcommand{\amc}{\texttt{amc}}

\newcommand{\fao}{\texttt{FA1.1}}
\newcommand{\fat}{\texttt{FA1.3}}
\newcommand{\faf}{\texttt{FA1.5}}
\newcommand{\fan}{\texttt{FA1.9}}
\newcommand{\cahp}{\texttt{CA-HP}}
\newcommand{\camp}{\texttt{CA-MP}}
\newcommand{\calp}{\texttt{CA-LP}}

\shorttitle{Fractal particles in the IM Lup disk}
\shortauthors{Tazaki et al.}

\begin{document}

\title{Fractal aggregates of sub-micron-sized grains in the young planet-forming disk around IM Lup}

\correspondingauthor{Ryo Tazaki}
\email{ryo.tazaki1205@gmail.com}

\author[0000-0003-1451-6836]{Ryo Tazaki}
\affil{Anton Pannekoek Institute for Astronomy, University of Amsterdam, Science Park 904, 1098XH Amsterdam, The Netherlands}
\affil{Institute of Planetology and Astrophysics, Université Grenoble Alpes, 38000 Grenoble, France}
\affil{Astronomical Institute, Graduate School of Science,
Tohoku University, 6-3 Aramaki, Aoba-ku, Sendai 980-8578, Japan}

\author[0000-0002-4438-1971]{Christian Ginski}
\affil{Leiden Observatory, Leiden University, P.O. Box 9513, 2300 RA Leiden, The Netherlands}

\author[0000-0002-3393-2459]{Carsten Dominik}
\affil{Anton Pannekoek Institute for Astronomy, University of Amsterdam, Science Park 904, 1098XH Amsterdam, The Netherlands}




\begin{abstract}
Despite rapidly growing disk observations, it remains a mystery what primordial dust aggregates look like and what the physical and chemical properties of their constituent grains (monomers) are in young planet-forming disks. 
Confrontation of models with observations to answer this mystery has been a notorious task because we have to abandon a commonly used assumption, perfectly spherical grains, and take into account particles with complex morphology. In this Letter, we present the first thorough comparison between near-infrared scattered light of the young planet-forming disk around IM Lup and the light-scattering properties of complex-shaped dust particles. The availability of scattering observations at multiple wavelengths and over a significant range of scattering angles allows for the first determination of the monomer size, fractal dimension, and size of dust aggregates in a planet-forming disk.
We show that the observations are best explained by fractal aggregates with a fractal dimension of 1.5 and a characteristic radius larger than $\sim2~\mu$m. 
We also determined the radius of the monomer to be $\sim200$ nm, and monomers much smaller than this size can be ruled out on the premise that the fractal dimension is less than 2. 
Furthermore, dust composition comprising amorphous carbon is found to be favorable to simultaneously account for the faint scattered light and the flared disk morphology.
Our results support that planet formation begins with fractal coagulation of sub-micron-sized grains. All the optical properties of complex dust particles computed in this study are publicly available.
\end{abstract}



%

\section{Introduction}
Collisional growth of dust aggregates by Brownian motion is the very first step in planet formation. Dust coagulation in this step results in the formation of an aggregate whose fractal dimension is less than 2 \citep{Kempf99, Blum00, Krause04, Paszun06}, where the fractal dimension $\df$ is defined by $m\propto \ac^{\df}$; $m$ and $\ac$ being the mass and the characteristic radius of an aggregate, respectively.
Fractal aggregates have been found in the comet 67P/Churyumov--Gerasimenko \citep{Bentley16, Mannel16}.
However, despite the anticipation that they form in young planet-forming disks, there has been no observational evidence that convincingly shows their presence in such disks. 

Another key issue that needs observational clarification is the size of constituent grains of an aggregate, called monomers. Since the radius of the monomer affects the impact strength of aggregates \citep{Dominik97, Wada09}, it is an important quantity directly affecting the collisional growth of aggregates. \citet{RT22} recently estimated the monomer radius based on the degree of polarization of optical/near-infrared (IR) scattered light of planet-forming disks and placed the upper limit at $0.4~\mu$m. However, a firm determination of the monomer radius requires a detailed comparison of models with disk observations for each object, which is still an untapped task.

IM Lup is a K5 Class II T Tauri star \citep{Alcala17} with the age of $\sim$1 Myr \citep{Mawet12, Avenhaus18}. The large planet-forming disk surrounding the star has been extensively studied at optical/near-IR and millimeter wavelengths \citep{Pinte08, Panic09, Cleeves16, Avenhaus18, Andrews18, Oberg21}. Based on optical/near-IR disk-scattered light, \citet{Pinte08} suggested the presence of fluffy aggregates and/or ice-mantled grains. However, the detailed dust properties remain largely unknown.

\begin{figure*}[t]
\begin{center}
\includegraphics[width=\linewidth]{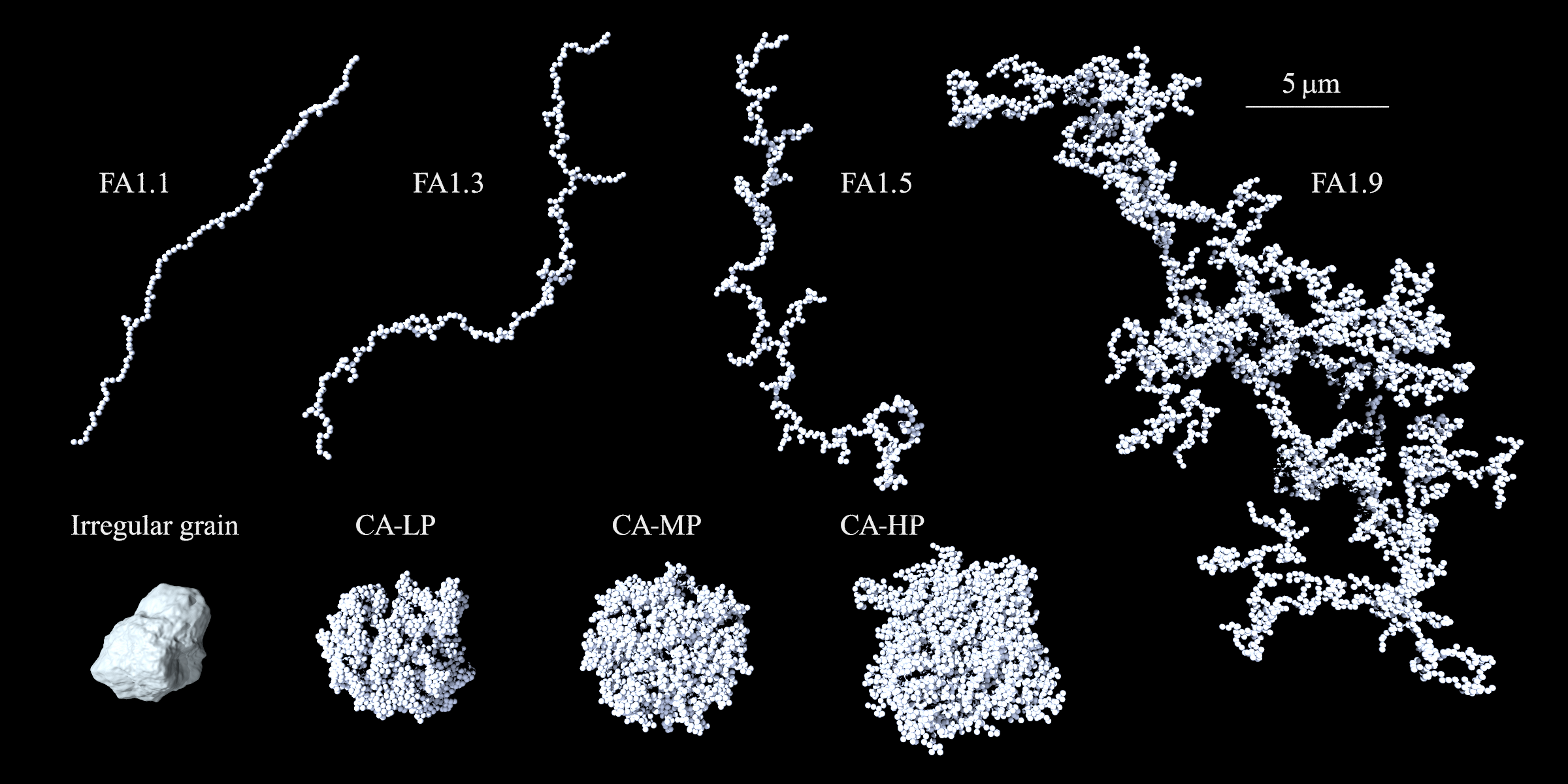}
\caption{Dust particle models used to model disk-scattered light around IM Lup: 
\fao~($N=128$, $\amon=100$ nm),
\fat~($N=256$, $\amon=100$ nm),
\faf~($N=512$, $\amon=100$ nm),
\fan~ ($N=4096$, $\amon=100$ nm), 
\cahp~($N=4096$, $\amon=100$ nm), 
\camp~($N=4096$, $\amon=100$ nm), \calp~($N=4096$, $\amon=100$ nm) and an irregular grain with $\av=1.6~\mu$m.}
\label{fig:particle}
\end{center}
\end{figure*}

Our aim in this Letter is therefore to unveil the detailed dust properties in the surface region of the IM Lup disk through near-IR polarimetric observations (Figure \ref{fig:obs}). To this end, we perform radiative transfer simulations of the disk based on a detailed numerical study of the optical properties of complex-shaped dust particles (Figure \ref{fig:particle}). The state-of-the-art approach enables us for the first time to show that dust particles in the IM Lup disk surface have already grown beyond a few microns in size, and they have a fractal structure with monomers of a radius of $\sim200$ nm. Our results support that planet formation begins with fractal coagulation of sub-micron-sized grains.

We also developed the \texttt{AggScatVIR} database\footnote{\texttt{AggScatVIR} repository \url{https://github.com/rtazaki1205/AggScatVIR}. v1.0.0: \url{https://doi.org/10.5281/zenodo.7547601}}, where all the optical properties calculated in this study and \citet{RT22} are publicly available. The database will be useful for retrieving the detailed dust properties in other planet-forming disks.

\section{Disk-scattered light around IM Lup} \label{sec:obs}
\begin{figure*}[t]
\begin{center}
\includegraphics[width=\linewidth]{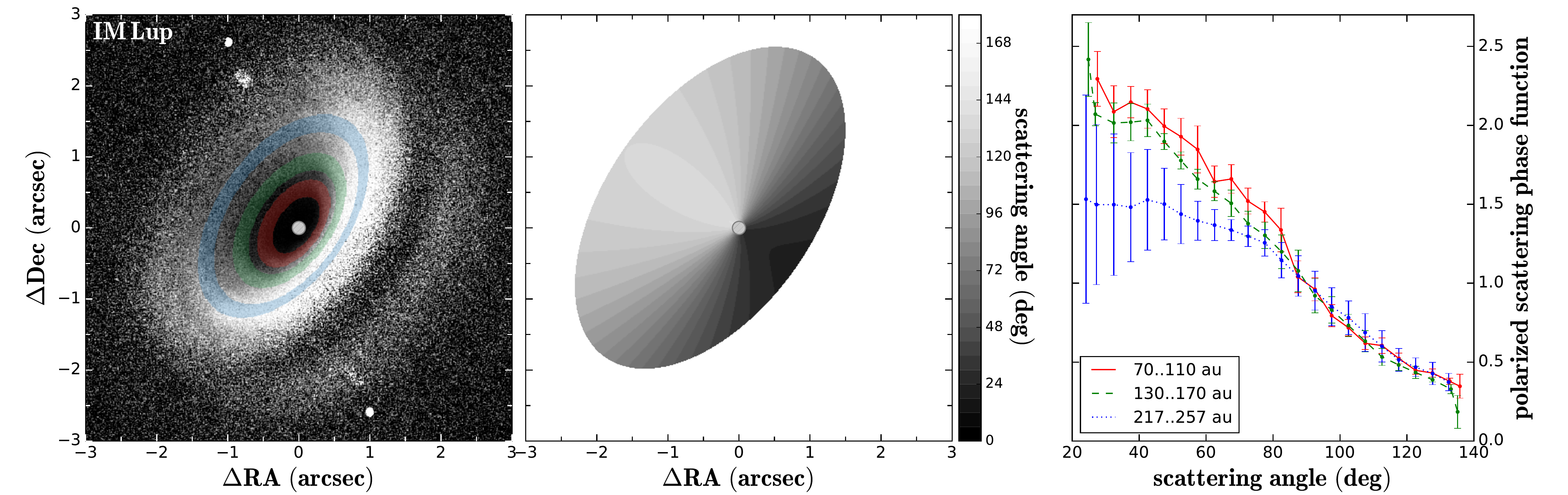}
\caption{Observed polarized scattered-light image for the IM Lup disk at the $H$ band (scaled with the square of the distance from the central star) overlaid with the extraction regions of the polarization phase function (left), the computed scattering angles at the disk surface (middle), and the extracted polarization phase functions (right). The phase functions were extracted at deprojected radii centered at $90$ au, $150$ au, and $237$ au with the 40 au width for each.}
\label{fig:obs}
\end{center}
\end{figure*}

Figure \ref{fig:obs} shows the observed near-IR polarized scattered-light image of the IM Lup disk taken by the Very Large Telescope (VLT)/Spectro-Polarimetric High-contrast Exoplanet REsearch (SPHERE) \citep{Avenhaus18}. The disk is visible in reflected light by dust particles in the disk surface. Each dust particle scatters the incoming unpolarized stellar light and produces linearly polarized light with an efficiency dependent on dust size, structure, and composition. Therefore, a detailed analysis of the polarized scattered light allows us to diagnose the properties of dust particles.

We focus on two observational quantities. The first quantity is {\it the disk polarization flux and its color}. \citet{Avenhaus18} measured the ratio of polarized disk flux to total flux (stellar and disk flux) to be $0.53\%\pm0.06\%$ and $0.66\%\pm0.05\%$ in the $J$ and $H$ bands, respectively. Their ratio gives $J/H=0.81\pm0.12$; thus, the disk polarization color is reddish ($J/H<1$). This result already suggests that dust particles are micron sized or even larger because the color would become bluish ($J/H>1$) otherwise due to Rayleigh scattering \citep{Bohren83, Mulders13}.

The second quantity is {\it the polarization phase function}, which describes the amount of polarized disk-scattered light at each scattering angle. 
This quantity is another key to narrowing down the properties of dust particles \citep{Ginski16, Stolker16, Milli19, Olofsson22, Engler22, Ginski23}. 
\citet{Ginski23} extracted the polarization phase functions of the IM Lup disk based on the VLT/SPHERE images at the $J$ and $H$ bands taken by \citet{Avenhaus18}. The phase functions were extracted at three different disk radii centered at 90 au, 150 au, and 237 au, as shown in Figure \ref{fig:obs} (see \citet{Ginski23} and Appendix \ref{sec:extract} for more detailed extraction procedures).
The polarization phase functions for 90 au and 150 au are similar to each other, whereas the one for 237 au deviates from the other two at a scattering angle below 80$^\circ$.
However, the one for 237 au has large error bars, and it is unclear if this deviation is real. 
Because of its larger errors, we do not focus on the phase function for 237 au. Also, given the similarity between the two, it is reasonable to assume that the 90 au and 150 au phase functions probe similar dust particles (see Appendix \ref{sec:sim} for more details). We will therefore focus on the polarization phase function at 90 au as representative of this disk.

It is worth bearing in mind that the polarization phase function extracted from a disk image does not necessarily coincide with the scattering matrix element of each dust particle (e.g., $-S_{12}$ in \cite{Bohren83}). This is because the disk is optically thick, and the emergent intensity could be affected by multiple scattering and limb brightening (see Appendix \ref{sec:appphase} for details). To fully account for these effects, we first create a model image using a radiative transfer calculation, then extract the phase function from the model image using the same technique as we did for the observed image, and compare it to the observation.

\begin{deluxetable*}{lcccccccccccccccc}
\tablecaption{Volume-equivalent and characteristic radii (normalized to monomer radius) and porosities of dust aggregates \label{tab:size}}
\tablewidth{\linewidth}
\tabletypesize{\scriptsize}
\tablehead{
\colhead{$N$} & \colhead{$\av/\amon$} & \multicolumn{7}{c}{Characteristic Radius $\ac/\amon$}  & \multicolumn{7}{c}{Porosity $\mathcal{P}=1-N(\amon/\ac)^3$ (\%)}  & \\ 
\cline{3-9}\cline{11-17}
\colhead{} & \colhead{} & \colhead{FA1.1} & \colhead{FA1.3} & \colhead{FA1.5} & \colhead{FA1.9} & \colhead{CA-HP} & \colhead{CA-MP} & \colhead{CA-LP}  & & \colhead{FA1.1} & \colhead{FA1.3} & \colhead{FA1.5} & \colhead{FA1.9} & \colhead{CA-HP} & \colhead{CA-MP} & \colhead{CA-LP}  
} 
\startdata
8 & 2.00 & 5.12 & 4.50 & 3.92 & 3.59 & 3.25 & 2.57 & 2.23 &&
94.03 & 91.21 & 86.71 & 82.74 & 76.73 & 52.91 & 27.78 \\
16 & 2.52 & 9.83 & 7.87 & 6.42 & 5.54 & 4.45 & 3.42 & 2.97 &&
98.31 & 96.72 & 93.96 & 90.57 & 81.82 & 60.12 & 38.68 \\
32 & 3.17 & 18.6 & 13.5 & 10.3 & 7.68 & 6.00 & 4.68 & 3.92 &&
99.50 & 98.71 & 97.09 & 92.94 & 85.17 & 68.87 & 46.73 \\
64 & 4.00 & 34.9 & 23.1 & 16.5 & 10.5 & 7.79 & 6.17 & 5.21 &&
99.85 & 99.48 & 98.56 & 94.55 & 86.47 & 72.78 & 54.86 \\
128 & 5.04 & 65.6 & 39.5 & 26.2 & 15.7 & 9.87 & 7.95 & 6.75 &&
99.95 & 99.79 & 99.29 & 96.68 & 86.68 & 74.57 & 58.37 \\
256 & 6.35 & 123.2 & 67.3 & 41.6 & 21.8 & 12.4 & 10.2 & 8.72 &&
99.99 & 99.92 & 99.64 & 97.53 & 86.55 & 75.93 & 61.41 \\
512 & 8.00 & \nodata & 114.7 & 66.0 & 32.4 & 15.6 & 12.9 & 11.2 &&
\nodata & 99.97 & 99.82 & 98.49 & 86.41 & 76.23 & 63.74 \\
1024 & 10.1 & \nodata & \nodata & 104.8 & 49.5 & 19.6 & 16.5 & 14.3 &&
\nodata & \nodata & 99.91 & 99.16 & 86.43 & 77.12 & 65.11 \\
2048 & 12.7 & \nodata & \nodata & \nodata & 67.2 & 24.8 & 20.9 & 18.3 &&
\nodata & \nodata & \nodata & 99.33 & 86.58 & 77.63 & 66.51 \\
4096 & 16.0 & \nodata & \nodata & \nodata & 100.4 & 31.4 & 26.4 & 23.4 &&
\nodata & \nodata & \nodata & 99.60 & 86.82 & 77.81 & 68.05 \\
\enddata
\end{deluxetable*}

\begin{deluxetable*}{lccccc}
\tablecaption{Number of monomers considered in this study \label{tab:mon}}
\tablewidth{0pt}
\tablehead{
\colhead{$\amon$} & \colhead{100 nm} & \colhead{150 nm} & \colhead{200 nm} & \colhead{300 nm} & \colhead{400 nm} 
}
\startdata
FA1.1 & 8,16,32,64,128,256 & 8,16,32,64 & 8,16,32,64 & 8,16,32 & 8,16 \\
FA1.3 & 8,16,32,64,128,256,512 & 8,16,32,64,128 & 8,16,32,64,128 & 8,16,32,64 & 8,16,32 \\
FA1.5 & 8,16,32,64,128,256,512,1024 & 8,16,32,64,128,256 & 8,16,32,64,128,256 & 8,16,32,64,128 & 8,16,32,64 \\
FA1.9 & 8,16,32,64,128,256,512,1024,2048,4096 & 8,16,32,64,128,256,512 & 8,16,32,64,128,256,512 & 8,16,32,64,128,256 & 8,16,32,64,128 \\
CA-HP & 8,16,32,64,128,256,512,1024,2048,4096 & \nodata & 8,16,32,64,128,256,512 & \nodata & 8,16,32,64 \\
CA-MP & 8,16,32,64,128,256,512,1024,2048,4096 & \nodata & 8,16,32,64,128,256,512 & \nodata & 8,16,32,64 \\
CA-LP & 8,16,32,64,128,256,512,1024,2048,4096 & \nodata & 8,16,32,64,128,256,512 & \nodata & 8,16,32,64 \\
\enddata
\end{deluxetable*}

\begin{table}[t]
\caption{Real ($n$) and imaginary ($k$) parts of the refractive index of the two mixtures.}
\label{tab:opcont}
\centering
\begin{tabular}{cccccc} 
\hline\hline          
Composition Model & \multicolumn{2}{c}{\org}  && \multicolumn{2}{c}{\amc}\\ 
\cline{2-3}
\cline{5-6}
$\lambda$ ($\mu$m) & $n$ & $k$ & & $n$ & $k$ \\
\hline
 3.78 & 1.53 & 0.0219 && 2.13 & 0.393 \\
 2.18 & 1.47 & 0.0134 && 1.98 & 0.385 \\
 1.63 & 1.48 & 0.0138 && 1.92 & 0.404 \\
 1.25 & 1.49 & 0.0104 && 1.86 & 0.420 \\
 1.04 & 1.49 & 0.0108 && 1.81 & 0.434 \\
0.735 & 1.50 & 0.0119 && 1.70 & 0.468 \\
0.554 & 1.51 & 0.0138 && 1.59 & 0.472 \\
\hline
\end{tabular}
\tablecomments{The material densities of the mixtures are 1.6487 g cm$^{-3}$ (\org) and 1.7779 g cm$^{-3}$ (\amc).}
\end{table}   

\section{Models and Methods}

To model the two observational quantities of the IM Lup disk, we perform 3D Monte Carlo radiative transfer simulations by using \texttt{RADMC-3D v2.0} \citep{Dullemond12}. We constructed a disk model for IM Lup so as to reproduce the observed disk geometry reported in \citet{Avenhaus18}. The disk model and the data reduction, including the phase function extraction from each model image, are described in Appendix \ref{sec:rtmodel}. 

In the radiative transfer simulations, we consider a diverse set of dust particle morphology, as shown in Figure \ref{fig:particle}.
In what follows, the term {\it particle} is used to loosely refer to either {\it grain} (a monolith-like solid) or {\it aggregate} (a cluster of grains). We consider three types of dust particles, and the methods to generate them are described in Appendix \ref{sec:dustgen}. The first type is solid irregular grains, which correspond to the case where no coagulation has been taking place in the disk surface.

The second type is fractal aggregates; they are labeled by \fao, \fat, \faf, \fan~in Figure \ref{fig:particle}. We consider four different fractal dimensions ranging from 1.1 to 1.9, which is indicated by the number that follows \texttt{FA}. These fractal aggregates are expected to be formed by primordial dust coagulation, such as the one driven by Brownian motion \citep{Kempf99, Blum00, Krause04, Paszun06}.

The third type is compact aggregates\footnote{Such compact aggregates are sometimes referred to as fractal aggregates with a fractal dimension close to three. In this Letter, the term "{\it fractal aggregates}" is only used for those having a fractal dimension less than two.}; they are labeled by \cahp, \camp, \calp~in Figure \ref{fig:particle}. The letters following CA represent the amount of porosity: high porosity (HP), moderate porosity (MP), and low porosity (LP). 
The detection of (sub)millimeter-wave scattering polarization for the IM Lup disk \citep{Hull18, Stephens20} suggests that dust particles are likely compact aggregates at least in the midplane \citep{RT19mm, BW21}, although dust particles in the disk surface may not necessarily be.

We assume monodisperse and spherical monomer grains with a radius of $\amon=100$, $200$, and $400$ nm for compact aggregates and $\amon=100$, $150$, $200$, $300$, and $400$ nm for fractal aggregates. The summary of the monomer and aggregate radii is given in Tables \ref{tab:size} and \ref{tab:mon}. 
Irregular grains and monomers are made of a mixture of water ice \citep{Warren08}, pyroxene silicate (Mg$_{0.7}$Fe$_{0.3}$SiO$_3$) \citep{Dor95}, carbonaceous material, and troilite \citep{Henning96} with the mass ratios proposed by \citet{Birnstiel18}. Since the actual form of carbonaceous material is highly uncertain, we consider two possibilities: organics \citep{Henning96} or amorphous carbon \citep{Zubko96}. The refractive indices of the mixtures are calculated by using the Bruggeman mixing rule, and the results are summarized in Table \ref{tab:opcont}.
To represent the monomer model, we will use the following format: \texttt{COMP}\texttt{XXX}, where \texttt{COMP} and \texttt{XXX} specify the monomer composition and the monomer radius in units of nanometers, respectively. \texttt{COMP} has either \org~or \amc, where the former and the latter correspond to the mixture containing organics and amorphous carbon, respectively. For example, \amc200 represents a monomer grain of a radius of 200 nm and is made of a mixture containing amorphous carbon. 

We compute the optical properties of dust particles by using the $T$-Matrix method \citep[][and references therein]{Mackowski96} for aggregates and discrete dipole approximation (DDA) for irregular grains \citep{Purcell73, Draine93}. 
Both techniques are known as exact numerical techniques to calculate the optical properties of nonspherical particles. For the $T$-Matrix calculations, we use \texttt{MSTM-v3.0} with analytical orientation averaging and four-realization averaging  for each model \citep{Mackowski11}. For the DDA calculations, we use \texttt{ADDA} \citep{Yurkin11} with averaging of 58 orientations and 10 realizations.

The obtained optical properties are then averaged by considering a particle-size distribution obeying 
\begin{equation}
n(\av)d\av \propto \av^{-3.5}d\av~(\amin\le \av\le \amax), 
\end{equation}
where $n(\av)d\av$ is the number density of particles within a radius range [$\av$, $\av+d\av$], $\av=(3V/4\pi)^{1/3}$ is the volume-equivalent radius; $V$ being its material volume, and $\amin$ and $\amax$ are the minimum and maximum volume-equivalent radii, respectively. The discrete sampling of $\av$ is shown in Table \ref{tab:size}.
The minimum size of the distribution is set as $\amin=2\amon$ for aggregates and $\amin=0.2~\mu$m for irregular grains. The maximum particle radius $\amax$ is a parameter of this study. Since the volume-equivalent radius is not a good indicator of the apparent size of an aggregate, we also use the characteristic radius $\ac$ \citep{Mukai92}, and let $N_\mathrm{max}$ and $\acmax$ denote the number of monomers and the characteristic radius of the maximum aggregate in the size distribution, respectively.

We investigated 360 sets of dust particle models: 20 irregular grain models (10 maximum grain radii; 2 compositions), 126 compact aggregate models (3 porosity models; 2 compositions; 3 monomer radii; on average 7 maximum aggregate radii), and 214 fractal aggregate models (4 fractal dimensions, 2 compositions, 5 monomer radii; on average 5.35 maximum aggregate radii).

\section{Results} \label{sec:result}

\begin{figure*}[t]
\begin{center}
\includegraphics[width=\linewidth]{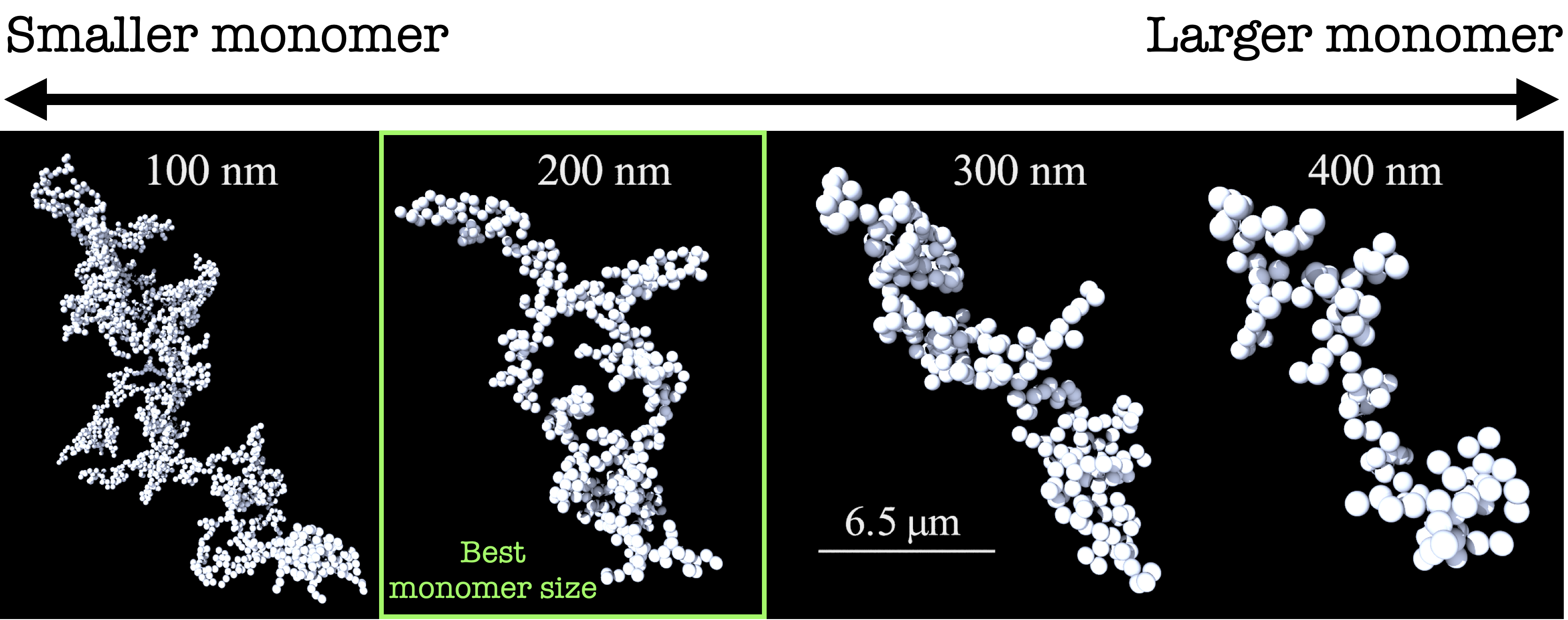}
\includegraphics[width=\linewidth]{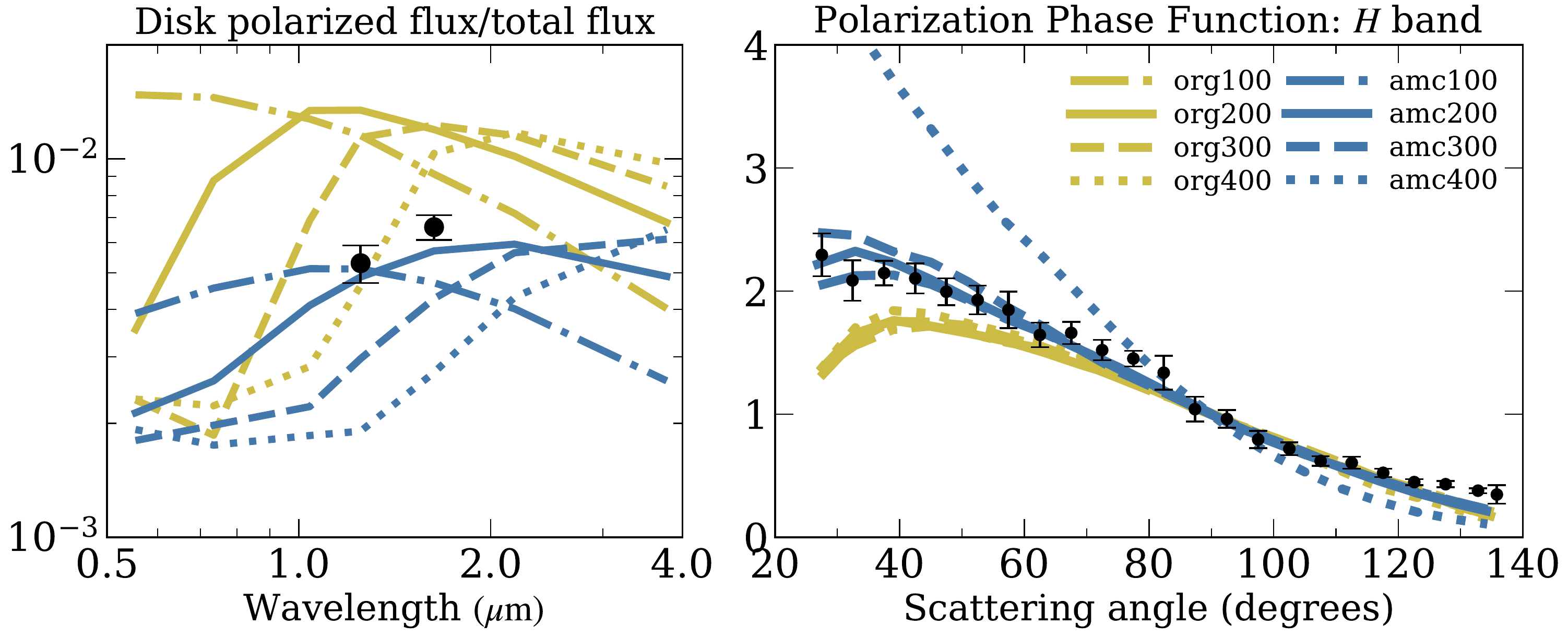}
\caption{Comparison of polarized scattered light of the IM Lup disk and simulated one for \fan~with $\acmax=6.5\pm0.22~\mu\mathrm{m}$ but consisting of different monomer radii and compositions. The top images show the largest aggregate in the size distribution of each model. Adopted aggregate parameters are ($N_\mathrm{max}$, $\amon$)=(2048, 100 nm) (dotted-dashed), (512, 200 nm) (solid), (256, 300 nm) (dashed), (128, 400 nm) (dotted). 
The bottom left and right panels show the polarized flux and the polarization phase function (normalized to $90^\circ$) at the $H$ band, respectively. Blue and yellow colors represent the \amc~and \org~composition models, respectively.}
\label{fig:mono}
\end{center}
\end{figure*}
\begin{figure*}[t]
\begin{center}
\includegraphics[width=\linewidth]{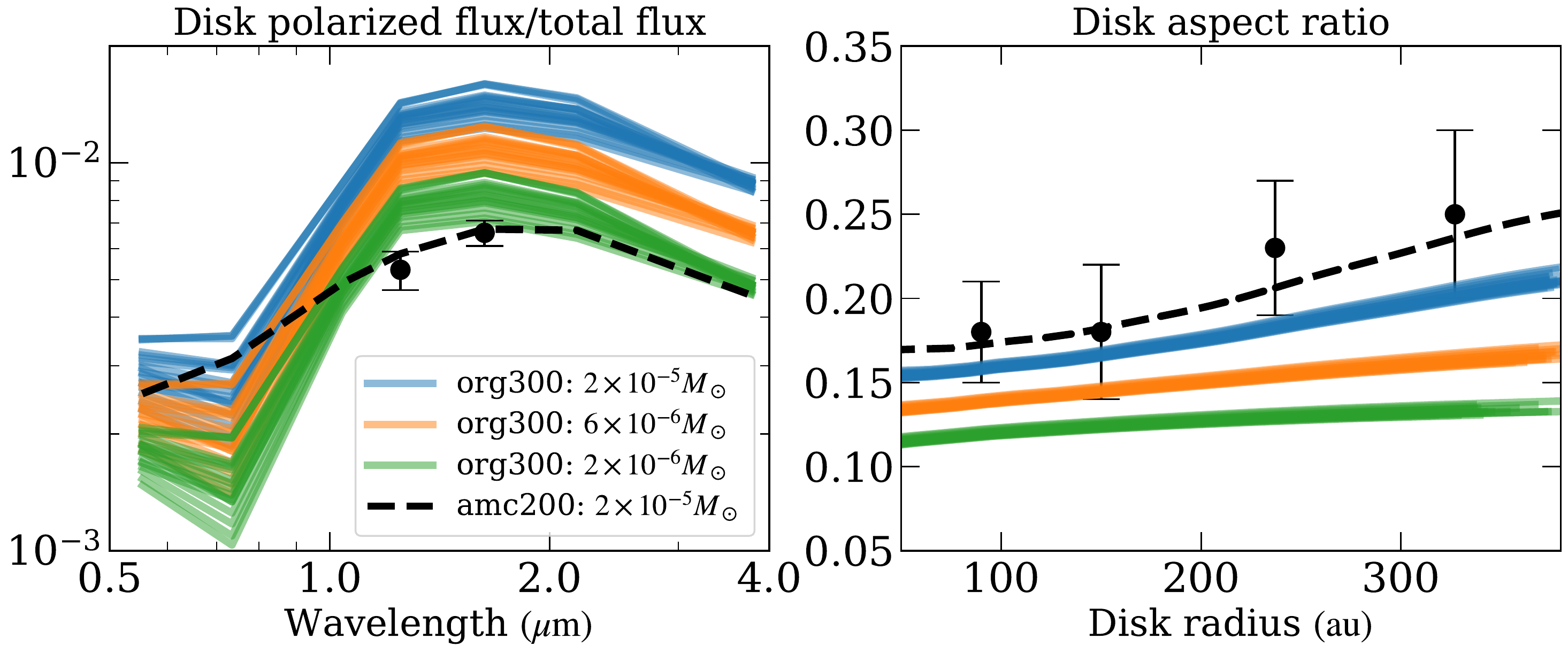}
\caption{Effect of dust disk mass on the polarized flux and the aspect ratio of the disk. The aspect ratio of each dust disk model was measured at a disk height where the optical depth measured from the star becomes unity. 
The blue, orange, and green lines represent the results for all fractal models with \org300 (18 models in total, see Table \ref{tab:mon}) with the dust disk mass $M_\mathrm{dust}=2\times10^{-5}M_\sun$ (fiducial), $6\times10^{-6}M_\sun$ and $2\times10^{-6}M_\sun$, respectively. The dust disk mass does not include the contribution of pebble-sized grains that settle into the midplane.
The black dashed line represents the results for the fiducial disk mass with \faf~with $N_\mathrm{max}=256$ with the monomer model of \amc200 (see Section \ref{sec:bestpart}). The observed faint scattered light and large aspect ratio of the disk surface point toward a highly absorbing composition of each monomer.}
\label{fig:diskmass}
\end{center}
\end{figure*}

\subsection{Disk polarization flux and color} \label{sec:imlupmono}

Figure \ref{fig:mono} compares the disk-scattered light properties for fractal aggregates having a similar structure and radius, but different monomer models (\fan~with $\acmax=6.5\pm0.22~\mu$m; see top panels in Figure \ref{fig:mono}). 
We found that the properties of monomers strongly affect the polarized scattered-light flux, from which we can constrain the monomer properties.

The composition of monomers has a significant impact on the absolute level of disk flux. The \org~models produce higher polarized fluxes than the \amc~models because the former composition has a higher scattering albedo at optical/near-IR wavelengths. 
If we reduce the disk mass, the disk flux could be reduced as well, as shown in Figure \ref{fig:diskmass}. However, such a reduction results in lowering the scattering surface, and these models fail to explain the aspect ratio of the disk regardless of aggregate size and fractal dimension. Moreover, we arrived at the same conclusion even considering compact aggregates. All compact aggregate models with the \org~composition fail to reconcile the disk flaring and polarized flux.
Therefore, given the relatively faint scattered light and its flared disk geometry of the IM Lup disk, a highly absorbing monomer composition (e.g., the \amc~composition) is favorable. 

The observed reddish disk color favors a not-too-small monomer size.
Figure \ref{fig:mono} (bottom left) shows that models with $\amon=100$ nm produce bluish dependencies ($J/H>1$) regardless of composition and are inconsistent with the observation. On the premise that $\df<2$, which we think is likely (see Sections \ref{sec:bestpart}), the presence of even smaller monomers can be ruled out by the following argument. A {\it total-intensity} disk color turned out to be nearly gray for all models shown in Figure \ref{fig:mono}. This tendency remains unchanged no matter how large the aggregate is unless we consider compact aggregates \citep{RT19ir}. The observed reddish-polarization color thereby needs to be explained by a decrease in the degree of polarization for shorter wavelengths. The degree of polarization of an aggregate is predominantly determined by the properties of each monomer \citep{RT16, RT22}. Therefore, the degree of polarization of {\it each monomer} has to decrease for shorter wavelengths. Since such a property is never achieved by Rayleigh-scattering monomers such that $\xmon=2\pi\amon/\lambda\ll1$, we can rule out $\amon\ll \lambda/2\pi\sim 200$ nm, where we substituted $\lambda=1.25~\mu$m (the $J$ band).

Contrary to the case of polarized flux, the impact of the monomer model on the polarization phase function is minor (Figure \ref{fig:mono} bottom right). 
Nevertheless, we can exclude some of the monomer models from it. For example, the \amc400 model gives rise to a too-steep phase function. 
Also, the curves for the aggregates with the \org~composition show a turnaround at a scattering angle of $\sim40^\circ$, while it is absent for the cases of \amc. This turnaround is caused by multiple scattering in the optically thick disk surface (see Figure \ref{fig:multiplescat} in Appendix \ref{sec:appphase}). The absence of the turnaround in the observed data also favors the \amc~composition. 

\subsection{Polarization phase function} \label{sec:bestpart}
\begin{figure*}[t]
\begin{center}
\includegraphics[width=\linewidth]{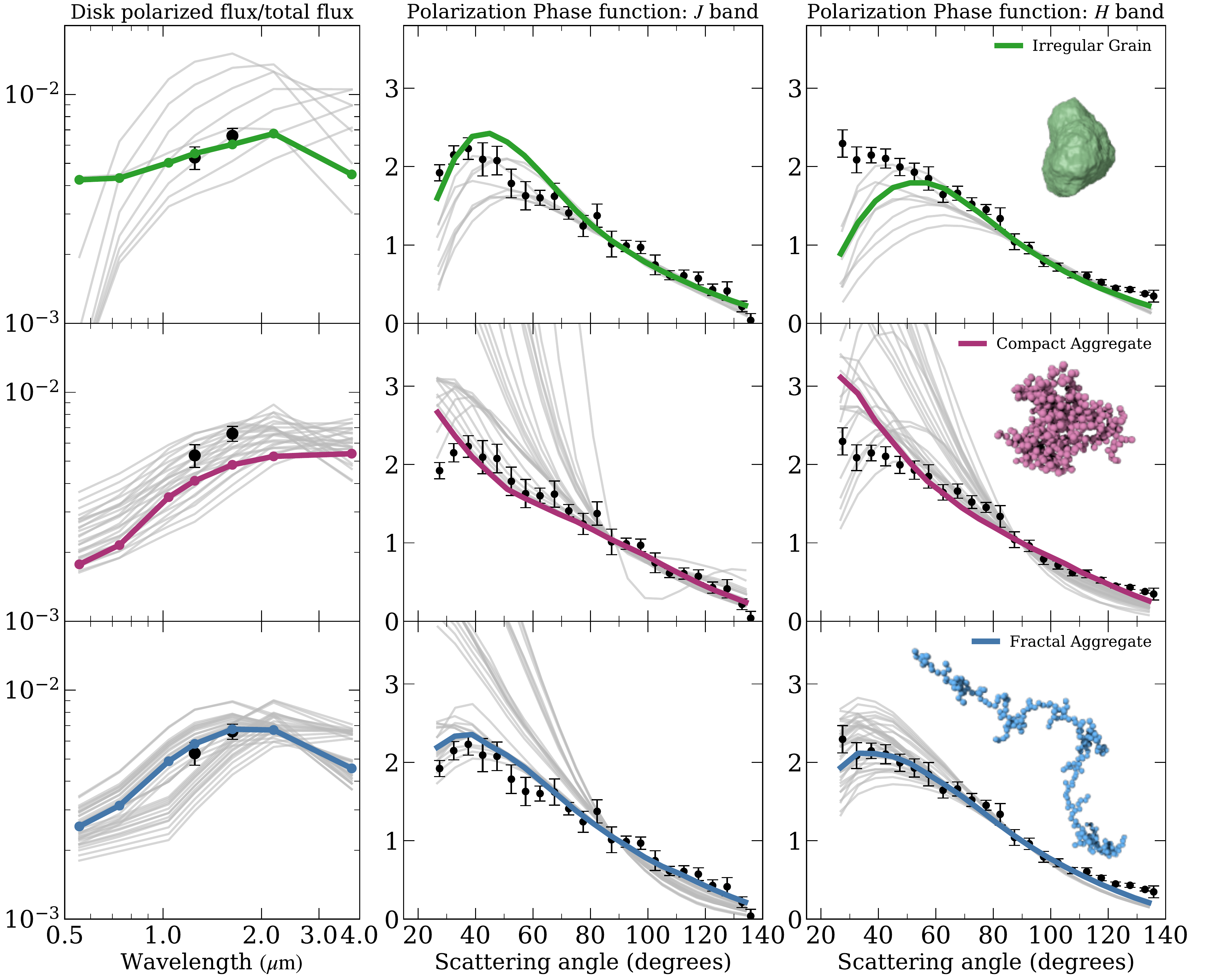}
\caption{Best dust models for the polarization phase function of the IM Lup disk. The top, middle, and bottom panels represent the results for irregular grains, compact aggregates, and fractal aggregates, respectively. From left to right panels, the disk-polarized flux and the polarization phase functions (normalized to a scattering angle of 90$^\circ$) at the $J$ and $H$ bands are shown. The thick lines represent the best fit among each particle type: the best irregular grain model ($\amax=0.504~\mu\mathrm{m}$ with \texttt{amc}: a reduced $\chi^2=6.8$), the best compact aggregate model (\cahp~with $N_\mathrm{max}=512$ and \texttt{amc200}: a reduced $\chi^2=5.0$), the best fractal model (\faf~with $N_\mathrm{max}=256$ and \texttt{amc200}: a reduced $\chi^2=3.6$). Thin gray lines in each panel are models that are consistent with the disk polarization color ($J/H=0.81\pm0.12$).}
\label{fig:survey}
\end{center}
\end{figure*}

Figure \ref{fig:survey} shows the polarization phase functions of various particle models, from which we can constrain the aggregation models. 
In the plots, we excluded models that have an inconsistent disk color, which basically resulted in rejecting large fractal aggregates with small monomers and small particles obeying Rayleigh scattering. For fractal and compact aggregates, we also excluded the \org~composition and \amc400, as these parameters are already shown to be inconsistent with the observation (Section \ref{sec:imlupmono}). 
As a result, the number of models plotted is 9, 27, and 36 for irregular grains, compact aggregates, and fractal aggregates, respectively. For each model, we assess the quality of fit by calculating a reduced $\chi^2$ including both data points of the $J$- and $H$-band phase functions. 

The best-fit model among irregular grains is $\amax=0.504~\mu\mathrm{m}$ with \texttt{amc}, which yields a reduced $\chi^2=6.8$. This model shows a rapid drop in polarized intensity at a scattering angle below $\sim50^\circ$ at the $H$ band and is inconsistent with the observation. The drop stems from two competing functions overlapping in the polarization phase function: one is the total-intensity phase function, which peaks at smaller scattering angles, and the other one is the degree of polarization, which decreases at these angles. For all irregular grain models, the latter function is dominant.

Polarized forward scattering by compact aggregates is much stronger than that of irregular grains. This is because the aggregates exhibit stronger forward scattering in total intensity, as they have larger sizes than irregular grains with the same mass. 
We also found that there is a large variation in polarization phase functions (see gray lines), and most of them overestimate the observed polarized forward scattering amplitude. The best-fit model among compact aggregates is \cahp~with $N_\mathrm{max}=512$ and \texttt{amc200} ($\amax=1.6~\mu$m and $\acmax=3.1~\mu$m), which yields a reduced $\chi^2=5.0$. However, this model underestimates the polarized flux. Although it could be enhanced by increasing the disk mass and then scattering the surface, the geometrically thicker disk would make the phase function even steeper (see Appendix \ref{sec:appphase}). 

Finally, we found that fractal models outperform the other two types of particle models. The best-fit model is \faf~with $N_\mathrm{max}=32, 64, 128, 256$ and \texttt{amc200}. All of them yield a reduced $\chi^2=3.6$ and are therefore the best fits among all particle models. These aggregates have sizes of $\amax=0.63$--$1.3~\mu$m and $\acmax=2.1$--$8.3~\mu$m. 
The upper bound of the aggregate size is ill-constrained as it corresponds to the larger end of the parameter range we studied. We think $\acmax>8.3~\mu$m is also a possible solution because the aggregate radius is insensitive to the results. These models can successfully reproduce the observed disk-polarized flux, its color, disk aspect ratio, and polarization phase functions at the two bands simultaneously (see also the dashed line in Figure \ref{fig:diskmass}). In what follows, we select the largest one (\faf~with $N_\mathrm{max}=256$ and \amc200) as representative of the best fractal models.

The polarization phase functions of the fractal models tend to be shallower than those of the compact aggregates, despite the fact that the fractal aggregates are much larger than the compact ones. This is due to a different number density of monomers. In a compact aggregate, the monomers are packed quite closely, and hence, scattered waves from the monomers are efficiently amplified by constructive interference. In a fractal aggregate, each monomer is relatively isolated so that fewer monomer pairs will be involved with constructive interference. As a result, forward scattering is less prominent in fractal models.

We also found, more importantly, that once fractal aggregates become micron sized, the quality of fit is insensitive to the aggregate radius and weakly dependent on the fractal dimension. 
As long as the monomer model is \amc200, there are a number of models that fit the observations almost equally well: \fao~with $\acmax\ge3.7~\mu$m ($\chi^2=3.9-4.1$), \fat~with $\acmax\ge2.7~\mu$m ($\chi^2=4.1-4.5$), \faf~with $\acmax\ge2.1~\mu$m ($\chi^2=3.6$), and \fan~with $\acmax\ge1.1~\mu$m ($\chi^2=4.0-5.0$). 
This explains why there is less variation in gray lines compared to compact aggregate models. This property is a direct consequence of the suppression of multiple scattering, that is, monomer-monomer electromagnetic interaction, due to its highly fluffy structure \citep{RT16}. Without multiple scattering, two scattered waves emanating from a monomer pair that is separated by much more than the wavelength are independent (incoherent) of each other. Scattered waves from monomers in a close neighborhood can still provide coherent scattering, and it is this component that dominantly contributes to scattered-light intensity. As a result, except for a scattering angle very close to $0^\circ$, the properties of scattered light will only be determined by what a local structure within an aggregate looks like and not by how large the aggregate is \citep{Berry86, RT19ir}. 

There are some outliers in the $J$-band phase functions of fractal models. They correspond to aggregates with \amc300. Thus, the monomer radius of 300 nm is unfavorable as well.

Although the difference in the reduced $\chi^2$ values between the best compact and fractal aggregate models is not significant, the fractal model is favorable because of its robust fitting behavior. 
Since the phase functions of compact aggregate models are sensitive to aggregate parameters, any fit requires significant fine-tuning. In contrast, those of fractal aggregates are very similar to each other, and most models that have a consistent disk color come close to the observed phase functions. Such properties allow us to model the observed scattered light without performing fine-tuning. 

\section{Discussion} 

\subsection{Are we observing primordial dust coagulation?} \label{sec:origin}

Our best-fit model suggests the presence of fractal aggregates with $\df=1.5$. These low-$\df$ aggregates naturally form via dust coagulation driven by Brownian motion \citep{Kempf99, Blum00, Krause04, Paszun06}, which is supposed to occur at the earliest step in planet formation. For example, \citet{Paszun06} obtained $\df=1.46$ for the Brownian coagulation in the ballistic collision limit. Here we discuss to what extent such aggregates can grow and survive in the disk surface after the disk age of $\sim1$ Myr.

Growth of aggregates by Brownian motion as a function of time $t$ can be approximated by \citep{Blum04, Krause04}
\begin{equation}
    N(t)=\left[(1-\gamma)\left(a\frac{t}{\tau_0}+c\right)\right]^{1/(1-\gamma)},\label{eq:size}
\end{equation}
where $\tau_0=(n_0\sigma_0 v_0)^{-1}$; $n_0$ is the number density of the monomers, $\sigma_0=4\pi \amon^2$ is the collision cross section, and $v_0=4\sqrt{kT/\pi \mmon}$ is the relative velocity between them, $\mmon$ is the mass of each monomer, $k$ is the Boltzmann constant, and $T$ is the gas temperature. We adopt $a=1.28$, $c=2.05$, $1/(1-\gamma)=1.71$, as suggested by microgravity experiments \citep{Krause04}.

To estimate the size of aggregates using Equation (\ref{eq:size}), we assume the \amc200 model, which is the best monomer model found in Section \ref{sec:imlupmono}. The number density of monomers $n_0$ is estimated by using the gas density model of the IM Lup disk in \citet{Zhang21} with a dust-to-gas ratio of 0.01 (after scaling the new GAIA distance). We also assume a temperature of 25 K. 
By substituting these values into Equation (\ref{eq:size}), we found 
$N(t=1~\mathrm{Myr})\sim10^3$ and $6\times10$ at the 90-au and 150-au disk surface ($z\sim 2H_\mathrm{g}$). These sizes are larger than the smallest aggregate we need to explain the scattered light ($N_\mathrm{max}\simeq32$). Therefore, coagulation driven by Brownian motion is a viable mechanism for forming the inferred fractal aggregates.

Suppose the vertical mixing of dust particles is active. In that case, particles in the disk surface region could be a mixture of those formed in situ and those stirred up from the midplane; the latter may have a nonfractal structure. To estimate whether this is the case or not, we calculate the vertical stirring timescale \citep{Dullemond04}
\begin{equation}
    t_\mathrm{stir}=\frac{1}{\alpha\Omega}\frac{z^2}{H_\mathrm{g}^2},
\end{equation}
where $z$ is the disk height, $H_\mathrm{g}$ is the gas pressure scale height, $\Omega$ is the Kepler angular frequency, and $\alpha$ is a nondimensional parameter for the vertical diffusion coefficient. 
With the gas scale height model of \citet{Zhang21}, the scattering surface derived in \citet{Avenhaus18} corresponds to $z\sim2 H_\mathrm{g}$. Since a typical disk model has a scattering surface at $3$--$4H_\mathrm{g}$, the dust disk around IM Lup is relatively flat, as also argued in \citet{Rich21}. 
At $r=150$ au, we have $\Omega^{-1}\approx300~\mathrm{yr}$. 
In order to have mixing during the age of the disk, we need $t_\mathrm{stir}<t_\mathrm{age}$, leading to $\alpha> (z/H_\mathrm{g})^2/(t_\mathrm{age}\Omega)\approx10^{-3}(t_\mathrm{age}/1~\mathrm{Myr})^{-1}$. However, there has been no evidence suggesting such a level of turbulence by molecular line observations toward the IM Lup disk.

There are some indirect measurements of the turbulent level in the IM Lup disk. \citet{Powell22} suggested $\alpha=1.5\times10^{-2}$ to explain the observed CO depletion within a timescale of $t_\mathrm{age}\approx1~\mathrm{Myr}$. However, a fast CO depletion within $\sim1~\mathrm{Myr}$ is feasible without invoking strong turbulence if the cosmic-ray ionization rate is as high as $10^{-16}$ with the help of other sequestration mechanisms \citep{Krijt20}. This rate is still a reasonable value at the outer surface region \citep{Indriolo15, Fujii22}. \citet{Franceschi22} estimated the $\alpha$ parameter to be $3\times10^{-3}$ to explain the near-IR disk thickness. Since they assumed compact spherical grains, the inferred $\alpha$ value could even be reduced by a factor of a few for fractal aggregates. Therefore, these indirect measurements do not necessarily require a turbulent level much larger than $10^{-3}$.

In a weakly turbulent disk ($\alpha<10^{-3}$), vertical settling must be slow enough to explain the observed disk thickness in IR scattered light. We then calculate the settling timescale defined by \citep{Dullemond04}
\begin{equation}
    t_\mathrm{sett}=\frac{4}{3\sqrt{2\pi}}\frac{A}{m}\frac{\Sigma_\mathrm{g}}{\Omega}\exp
    \left[-\frac{z^2}{2H_\mathrm{g}^2}\right], \label{eq:tsett}
\end{equation}
where  $A/m$ is the area-to-mass ratio of a fractal aggregate, and $\Sigma_\mathrm{g}$ is the gas surface density. 
At around the second ring ($r=150$ au), the gas surface density is $\Sigma_\mathrm{g}\approx 4$ g cm$^{-2}$ \citep{Zhang21}. 
A fractal aggregate of \faf~has an area-to-mass ratio of  $A/m\simeq0.7(A/m)_\mathrm{mono}$ at $N=256$, where $(A/m)_\mathrm{mono}$ is the area-to-mass ratio of each monomer grain \citep{RT21a}.
Thus, we have
\begin{equation}
    \frac{4}{3}\frac{A}{m}\Sigma_\mathrm{g}
    \approx8\times10^{4}\left(\frac{\rho_\mathrm{m}}{1.78~\mathrm{g~cm^{-3}}}\right)^{-1}\left(\frac{\amon}{200~\mathrm{nm}}\right)^{-1}.\label{eq:coup}
\end{equation}
By substituting Equation (\ref{eq:coup}) into (\ref{eq:tsett}), we obtain $t_\mathrm{sett}\approx 1$ Myr at $z=2H_\mathrm{g}$. The settling timescale at the scattering surface is thereby comparable to the age of the disk. The above estimate is also consistent with the findings of \citet{Verrios22}. Based on their hydrodynamic calculations without turbulence, the authors found that the model can reproduce the scattered-light morphology when the area-to-mass ratio of the smallest grains is $A/m=2.5\times10^4$ $\mathrm{cm}^2~\mathrm{g}^{-1}$, but not when $A/m=2.5\times10^3$ $\mathrm{cm}^2~\mathrm{g}^{-1}$ because settling occurs rapidly. Our best fractal aggregates have an area-to-mass ratio of $A/m\simeq1.5\times10^4$ $\mathrm{cm}^2~\mathrm{g}^{-1}$,  which agrees with their estimate within by a factor of two.

In summary, Brownian motion can form fractal aggregates with $\df=1.5$ large enough to explain the disk-scattered light. In the outer surface region of the disk, contamination of dust particles stirred up from the midplane would be limited unless the turbulent strength is $\alpha>10^{-3}$. Meanwhile, supported by the tight dynamical coupling of the fractal aggregates to gas, they settle down rather slowly to the disk midplane, which reasonably explains the observed disk thickness.

\subsection{The monomer radius for other planet-forming disks}
\begin{figure*}[t]
\begin{center}
\includegraphics[width=0.49\linewidth]{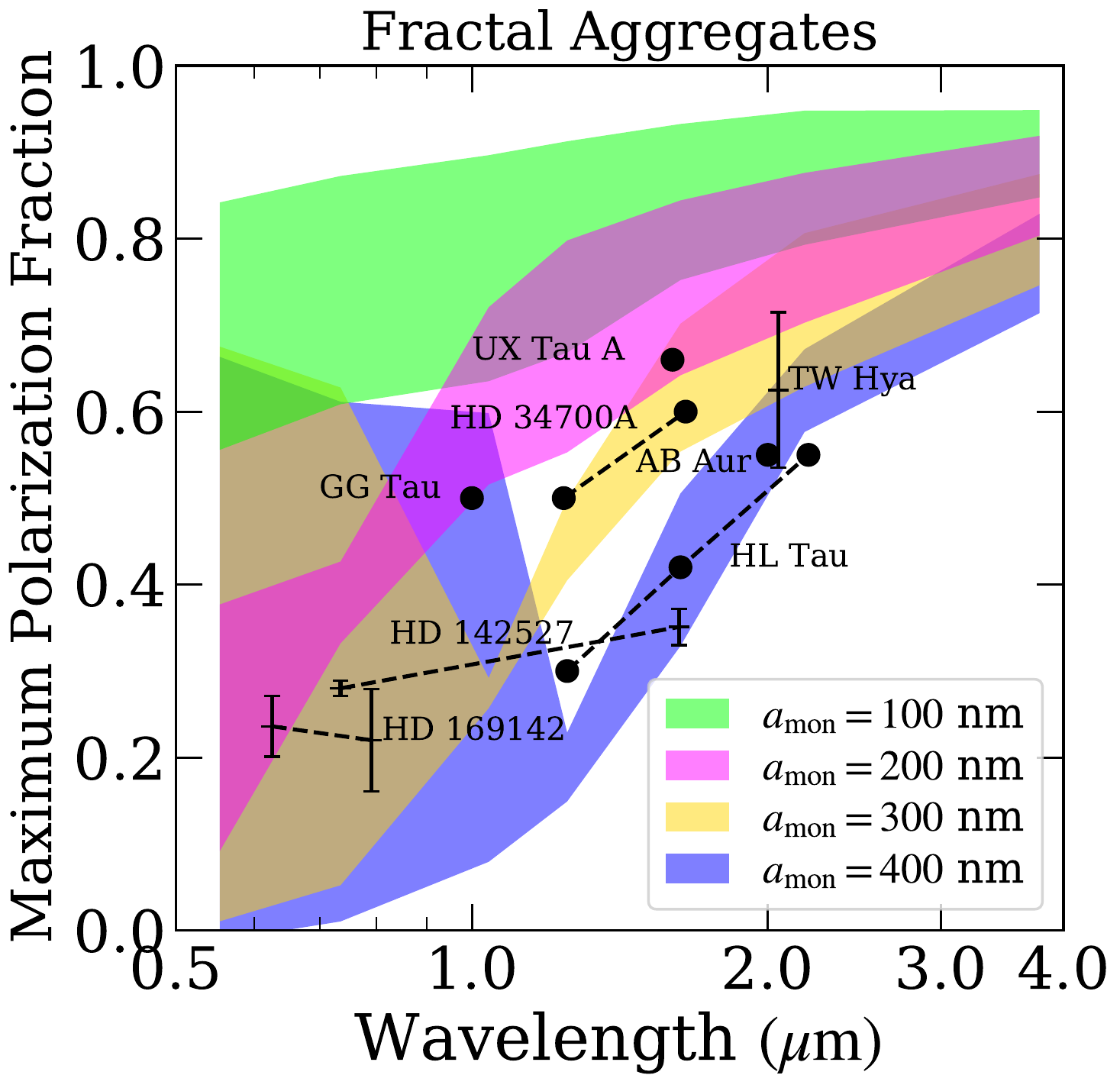}
\includegraphics[width=0.49\linewidth]{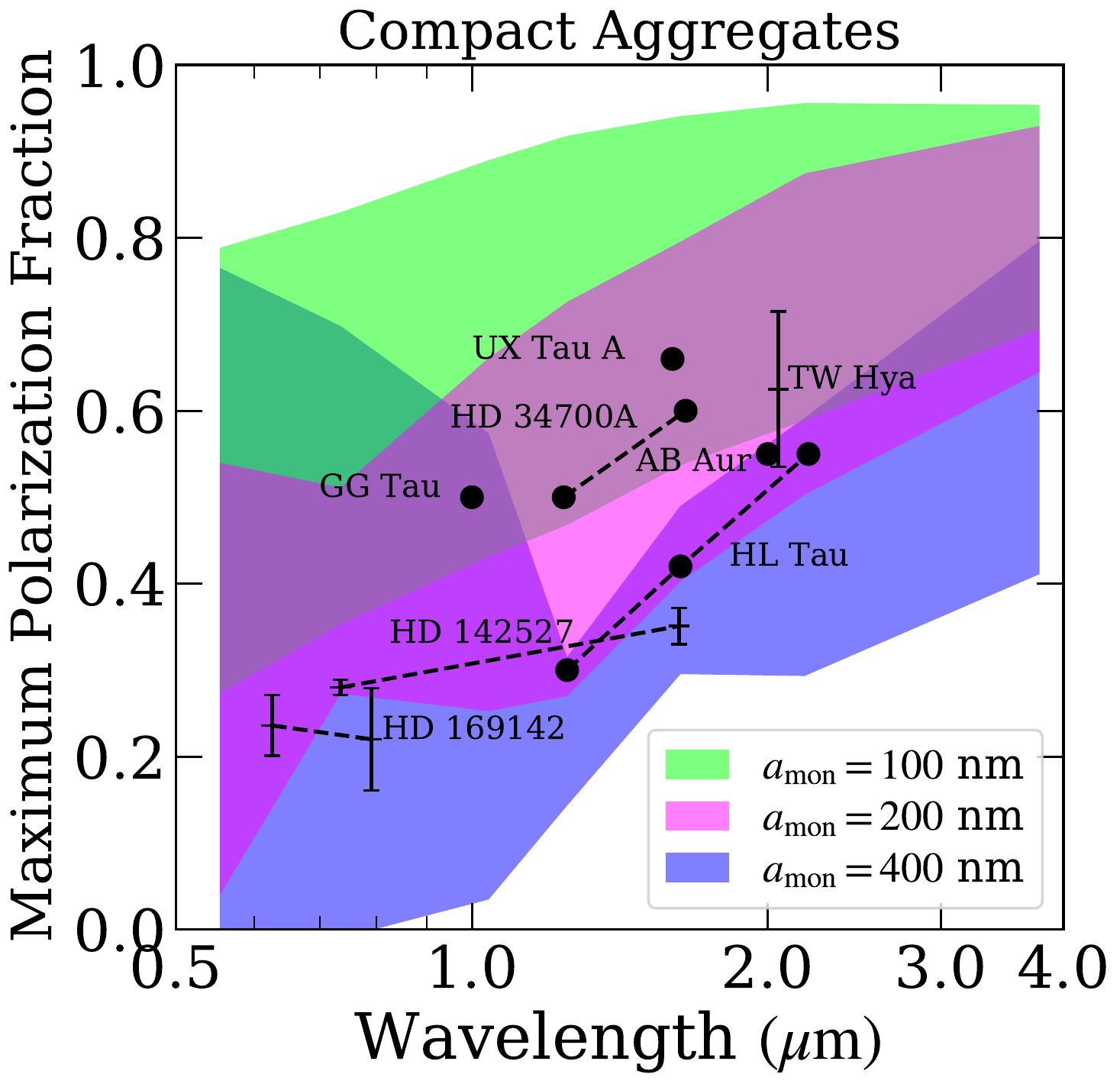}
\caption{Maximum degree of polarization of disk-scattered light for fractal aggregates (left) and compact aggregates (right). Each color represents the maximum polarization of aggregates with a specific monomer radius. For the left panel, the colored region shows the values of the maximum polarization for various aggregate radii, four fractal dimensions (\fao, \fat, \faf, \fan), and two monomer compositions (\org~and \amc). For the right panel, the colored region shows the values for various aggregate radii, three porosity models (\cahp, \camp, \calp), and two monomer compositions (\org~and \amc). For comparison, we overlaid the observed maximum polarization of various planet-forming disks. References: HD 142527 \citep{Hunziker21}, HD 169142 \citep{Tschudi21}, HD 34700 A \citep{Monnier19}, GG Tau \citep{Silber00}, AB Aur \citep{Perrin09}, UX Tau A \citep{Tanii12}, TW Hya \citep{Poteet18}, and HL Tau nebula region \citep{Murakawa08}.}
\label{fig:pmax}
\end{center}
\end{figure*}

\citet{RT22} argued that the degree of polarization is another key quantity for diagnosing the properties of the monomers. Figure \ref{fig:pmax} summarizes the maximum degree of polarization extracted from our radiative transfer simulations. Each colored region represents a range of the maximum degree of polarization that fractal or compact aggregates with a specific monomer radius can produce. In other words, it corresponds to the uncertainty range for various aggregate radii, composition, and porosity (fractal dimension).

As a general trend, a larger monomer radius leads to a lower degree of polarization.
A wider colored region in compact models is due to the degree of polarization being dependent on not only the monomer properties but also the aggregate parameters (porosity, aggregate size), whereas, in fractal models, it is predominantly determined by the monomer properties. The composition of monomers is another important factor for the degree of polarization. There is a tendency for an aggregate with \amc~to exhibit a higher degree of polarization than \org~because a higher scattering albedo of the latter composition facilitates multiple scattering at the disk atmosphere, which in turn reduces the degree of polarization \citep[e.g.,][]{Ma22}. The upper and lower bounds of the colored region, therefore, tend to be set by aggregates with the \amc~and \org~models, respectively.
The amount of amorphous carbon in each monomer also has a strong impact on optical polarization when $\amon=300$ and 400 nm, leading to a wider colored range for those models at optical wavelengths \citep{RT22}. 

The great advantage of using the maximum polarization as a diagnostic is that it is less sensitive to the disk structure and the inclination angle so that we can use it to infer the monomer properties in other disks.
In Figure \ref{fig:pmax}, we also plotted the maximum polarization fractions of several planet-forming disks, noting that the polarization fraction for the IM Lup disk has not yet been measured.
It turns out that the observed maximum polarization for the disks lies in the range of $\amon=100$--$400$ nm. Fractal models seem to favor a monomer radius of $200$--$400$ nm, whereas compact models favor $100$--$200$ nm. For HL Tau (nebula region), HD 163296, and HD 142527, the monomer radius of 100 nm appears to be unfavorable as none of the aggregate models can reproduce their relatively low degree of polarization. For HD 142527, the observed maximum polarization fraction is relatively low, and its wavelength dependence is shallow \citep{Hunziker21}. Such a tendency seems to be better explained by compact aggregates, supporting an earlier implication that aggregates are likely compact in HD 142527 \citep{RT21c, RT22}. 

To summarize, the monomer radius of other disks seems not to be too different from 200 nm, i.e. within by a factor of $\sim2$. We speculate that a monomer radius derived for the IM Lup disk, 200 nm, might be common among planet-forming disks at least for the outer surface regions.

\section{Conclusions}
We have shown that the near-IR polarized scattered-light observations of the IM Lup disk can be explained by fractal aggregates with a characteristic radius larger than $\sim2~\mu$m and a fractal dimension of 1.5. The monomer radius of $\sim200$ nm is favorable to explain the observed polarized scattered-light flux, and the monomer radius much smaller than 200 nm can be ruled out as long as the aggregates have a fractal dimension less than 2. Also, the faint polarized scattered light and its flared disk geometry suggest that each monomer is made of a highly absorbing composition (the \amc~model). These low-$\df$ fractal aggregates of sub-micron-sized monomers are shown to form by Brownian motion with a reasonable timescale in the IM Lup disk surface without suffering rapid settling. Our results support the idea that planet formation begins with fractal coagulation of sub-micron-sized grains via Brownian motion, as anticipated from laboratory experiments and numerical studies \citep{Kempf99, Blum00, Krause04, Paszun06}.

For the IM Lup disk, millimeter-wave scattering polarization has been detected \citep{Hull18, Stephens20}, which indicates the presence of less porous (sub)millimeter-sized aggregates \citep{RT19mm}. This result does not conflict with our conclusion because near-IR and millimeter wavelength observations probe different regions in the disk, and thereby different evolutionary stages of dust coagulation at least if perfect vertical mixing is not present. Combining these results, we speculate that dust coagulation initially proceeds in a fractal manner, but before reaching (sub)millimeter size, aggregates are compacted. In this way, multiwavelength disk scattered-light observations will shed light on the porosity evolution of dust aggregates in disks.

\acknowledgments
The authors thank the anonymous referee for his/her helpful comments.
R.T. acknowledges the JSPS overseas research fellowship. 
We thank Daniel Mackowski, Maxim A. Yurkin, and Cornelis Dullemond for making the MSTM, ADDA, RADMC-3D codes publicly available, respectively. We would also like to thank Bruce Draine for the availability of particle data of BA, BAM1, and BAM2, Yasuhiko Okada for providing a generation code for BCCA, and Tomas Stolker for providing the diskmap tool. R.T. also thanks useful discussion with Julien Milli and Daniel Price.
This work has made use of data from the European Space Agency (ESA) mission
{\it Gaia} (\url{https://www.cosmos.esa.int/gaia}), processed by the {\it Gaia}
Data Processing and Analysis Consortium (DPAC,
\url{https://www.cosmos.esa.int/web/gaia/dpac/consortium}). Funding for the DPAC has been provided by national institutions, in particular the institutions
participating in the {\it Gaia} Multilateral Agreement.

\software{\texttt{numpy} \citep{harris20}, \texttt{matplotlib} \citep{hunter07}, \texttt{RADMC-3D v2.0} \citep{Dullemond12} \texttt{MSTM v3.0} \citep{Mackowski11}, \texttt{ADDA} \citep{Yurkin11}, \texttt{diskmap} \citep{Stolker16}, \texttt{aggregate\_gen} \citep{Moteki19}}

\bibliography{cite}

\appendix

\section{Phase function extraction} \label{sec:extract}

The detailed extraction procedures are described in \citet{Ginski23}. Here we briefly summarize its procedure. For the extraction of the scattering phase functions in the $H$ and $J$ bands we fixed the inclination and position angle of the disk to the values given by \cite{Avenhaus18}, i.e. an inclination of 55$^\circ$ and a position angle of 325$^\circ$. We then used the power-law surface profile derived in the same study for the nominal extraction using the \texttt{diskmap} tool by \citet{Stolker16}.
They find a power-law (flaring) exponent of 1.271$\pm$0.197. 
We selected three extraction regions: $90$ au, $150$ au, $237$ au, corresponding to the surface ring locations seen in the scattered-light image. Each extraction region has a width of 40 au.
For the scaling factor of the power-law profile, we picked the direct measurement of the disk aspect ratio by \citet{Avenhaus18} at an angular separation of 0.96" corresponding to 149.6\,au (note that we rescaled the values given in \citet{Avenhaus18} with the new Gaia DR3 distance). Thus the nominal power-law profile for the surface is described by:

\begin{equation}
h_\mathrm{s}(r) = 0.046~\mathrm{au} \left(\frac{r}{1~\mathrm{au}}\right)^{1.27}.
\end{equation}

To estimate the uncertainty of the extracted phase functions due to the uncertainties of the surface profile we then repeated the extraction for two extreme cases, i.e. a flat and a strongly flared disk.
For the flat disk we used a flaring exponent of 1.07 (i.e. the lower bound given by the uncertainty interval), we moved the reference point outward to 154.3 au and increased the scaling factor accordingly to 0.095 au (in accordance with the uncertainty interval given by \citealt{Avenhaus18}). Conversely, for the flared disk we used a flaring exponent of 1.47, a reference point at 144.9 au, and a resulting scaling factor of 0.022 au. The uncertainties that we give for the final extracted phase function are then a (quadratic) combination of the range of the phase function given by the three different extractions for each angular bin and the statistical standard error of the pixel values in each angular bin for the nominal extraction (which captures the inherent noise in the image). This uncertainty is strongly dominated by the first term, i.e. how well we can constrain the surface profile determines to a large extent the precision with which we can extract the phase function.

\section{Simulation Methods} 
\subsection{Radiative transfer calculations and data reduction} \label{sec:rtmodel}

\begin{figure*}[t]
\begin{center}
\includegraphics[width=\linewidth]{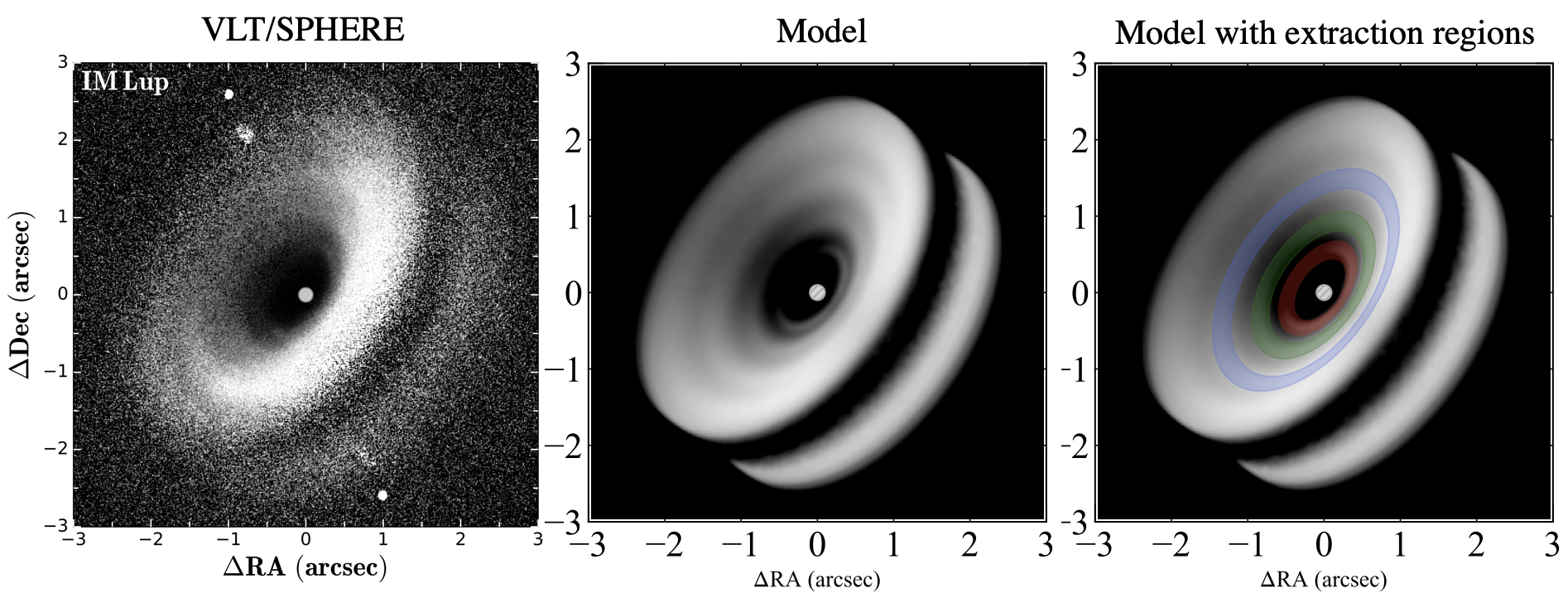}
\caption{Comparison of the observed disk image taken by VLT/SPHERE (left) and model images at the $H$ band without/with the phase function extraction regions (middle and right). The color scale shows the azimuthal polarization component $Q_\phi$ after correcting the drop-off of stellar light with distance. The assumed dust model is \faf~with $N_\mathrm{max}=256$ with the monomer model of \amc200, which is the best particle model derived in this study (Section \ref{sec:bestpart}).}
\label{fig:diskimage}
\end{center}
\end{figure*}

We summarize the star and disk models used in our radiative transfer simulations.
The stellar effective temperature and luminosity are set as 4266 K and 2.57 $L_\sun$, respectively \citep{Oberg21, Zhang21}. 
The central star is surrounded by an axisymmetric dust disk with a power-law surface density profile $\Sigma_\mathrm{d}(r)\propto r^{-1}$, where $\Sigma_\mathrm{d}(r)$ is the dust surface density at a radial distance of $r$. The inner and outer disk radii are 1 and 385 au, respectively. 
The inner radius is chosen well inside the coronagraphic mask and the outer radius coincides with the observed disk size (Figure \ref{fig:diskimage}). 
The total mass of the dust disk is set to $2\times10^{-5}M_\sun$ in a fiducial case, which corresponds to the total mass of the small-grain population suggested by \citet{Zhang21}. 
We ignored the large-grain (pebble) population, which dominates the total dust mass in the disk \citep{Zhang21, Franceschi22}, because these large grains reside close to the midplane and do not affect scattered-light images.

The vertical dust density profile is assumed to obey the Gaussian distribution $\rho_d(r,z)=\Sigma_\mathrm{d}(r)/\sqrt{2\pi h_\mathrm{d}^2}\exp[-z^2/2h_\mathrm{d}^2]$, where $z$ is the vertical height, and $h_\mathrm{d}$ is the dust scale height. 
We determine the scale height so as to reproduce the observational constraints by \citet{Avenhaus18}.
First, we first set a base height profile by $h_0(r)=5.1~\mathrm{au}(r/r_0)^{\beta}$, where $r_0=100~\mathrm{au}$, and $\beta=1.271$ and $1.6$ for $r<r_0$ and $r\ge r_0$, respectively.
The latter high $\beta$ value was introduced to better reproduce the large aspect ratios at the outer region \citep{Avenhaus18}.
We then added four axisymmetric rings by modifying the base scale height: 
\begin{equation}
h_\mathrm{d}=h_0(r)\left(1+\sum_iA_i\exp[-(r-r_i)^2/w_i^2]\right), \label{eq:hd}
\end{equation}
where $r_i$, $A_i$, and $w_i$ are the radius, amplitude, and width of the $i$-th ring. In this study, we adopt $r_1=90$ au, $r_2=150$ au, $r_3=237$ au, $r_4=327$ au, $A_1=0.1$, $A_2=0.05$, $A_3=0.1$, $A_4=0.15$, and $w_i=0.15r_i$. The assumed radial distances of the rings are taken from \citet{Avenhaus18} after correcting for an updated distance of $155.8$ pc. 
The observed image shows a relatively faint region inside 90 au \citep{Avenhaus18}. The faint region is presumably due to a shadow rather than a cavity because in the latter case, we would see a bright inner wall of the outer disk directly illuminated by the star. We are agnostic on the origin of this shadow-causing feature, as it is likely hidden inside the coronagraphic mask. Here we simply model it as a puffed-up ring with the parameters of $r_0=7$ au, $w_0=0.5$ au, and $A_0=1.2$ in Equation (\ref{eq:hd}). 

Given star and disk models, we perform scattering Monte Carlo simulations with a full scattering polarization treatment using \texttt{RADMC-3D v2.0} \citep{Dullemond12}. We use the spherical coordinate system. The zenith and azimuthal angles are linearly spaced within the range of [$\pi/3$,$2\pi/3$], and $[0,2\pi]$ with 512 and 256 grids. The radial grids are logarithmically equispaced with $512$ grids within the range of $[0.1~\mathrm{au}, 420~\mathrm{au}]$. Since we are interested in the outer disk region, where thermal emission hardly contributes to the observed intensity, we neglected the thermal emission of dust particles. We use $10^8$ photon packages for each scattering Monte Carlo simulation. We also assumed that dust particles of different sizes are well mixed throughout the disk. The obtained images are then convolved with a two-dimensional Gaussian function with an FWHM of 0.05\arcsec, which agrees with the size of the point-spread function of the observational data. Finally, we set all intensities to zero inside 0.1\arcsec~ around the star to mimic the coronagraphic mask. 

Our model can reproduce the overall properties of the observed disk (Figure \ref{fig:diskimage}). To see if our disk model reproduces the observed disk thickness, we calculated the disk surface height at which the optical depth measured from the star becomes unity for \faf~with $N_\mathrm{max}=256$ and the monomer model of \amc200. We obtained the aspect ratios of $0.17$, $0.18$, $0.21$, $0.24$ at 90 au, 150 au, 237 au, 327 au, respectively. These values are in agreement with the observations: $0.18\pm0.03$, $0.18\pm0.04$, $0.23\pm0.04$, $0.25\pm0.05$ at the same radii (see Figure \ref{fig:diskmass}). Once we obtained a model image, we integrated all polarized intensity of disk-scattered light outside the coronagraph to measure the disk-polarized flux. We then extracted the polarization phase functions using a model $Q_\phi$ image (after correcting the drop-off of stellar light with distance) at $90\pm20$ au, $150\pm20$ au, and $237\pm20$ au using the same method (Appendix \ref{sec:extract}), but with the correct disk heights calculated directly in each dust disk model and each observing wavelength.

\subsection{Generation of dust particle models} \label{sec:dustgen}

We generated \fan~ by means of Ballistic Cluster-Cluster Aggregation (BCCA) \citep[e.g.,][]{Mukai92}, which resulted in forming aggregates with a fractal dimension of 1.9. To generate aggregates with lower fractal dimensions, we adopt the sequential cluster-cluster aggregation algorithm developed in \citet{Filippov00}.
In this algorithm, aggregates are generated so that the resultant aggregate satisfies the fractal scaling law $N=\kf (\ag/\amon)^{\df}$, where $N$ is the number of monomers, $\amon$ is the radius of the monomer, $\ag$ is the radius of gyration, $\df$ is the fractal dimension, and $\kf$ is the fractal prefactor. 
Although \citet{Filippov00} also proposed another algorithm called the sequential particle-cluster aggregation, we avoid using this one because it leads to systematic errors in the two-point correlation function \citep{Filippov00, Krz14}, and the errors in the correlation function directly lead to systematic errors in the light-scattering properties.
For this reason, we prefer the sequential cluster-cluster aggregation algorithm for producing aggregates with $\df<2$. We assumed a fractal prefactor of $1.7$, $1.5$, $1.4$ for $\df=1.1$, $1.3$, and $1.5$, respectively \citep{RT21a}. To generate sequential CCA, we used a publicly available code \texttt{aggregate\_gen} \citep{Moteki19}. 
\cahp, \camp, and \calp~are generated by the BPCA (Ballistic Particle Cluster Aggregation), BAM1, and BAM2 algorithms \citep{Shen08} \footnote{The particle position data is taken from B. T. Draine's Web site
\url{https://www.astro.princeton.edu/~draine/agglom.html}}.
For all dust aggregates, we considered four realizations.
Lastly, solid irregular grains are generated by using the Gaussian random sphere (GRS) technique \citep{Muinonen96, Nousiainen03}. We assumed a power-law autocorrelation function with the index of $\nu=3.4$ and the relative standard deviation of radius $\sigma=0.2$ \citep[see][for the definition]{Nousiainen03}. For irregular grains, we considered ten realizations.
In DDA, we need to discretize each irregular grain into an array of dipoles. The total number of dipoles used in our DDA calculations is determined so as to satisfy $|m|kd\lesssim0.5$, where $d$ is the interdipole distance, $k$ is the wavenumber in a vacuum, and $m$ is the complex refractive index. 
For example, the average number of dipoles for irregular grains with $\av=1.6~\mu$m is 945186, which gives $|m|kd\simeq0.49$ at $\lambda=0.554~\mu$m (the shortest wavelength we studied) and $|m|kd\simeq0.20$ at $\lambda=1.630~\mu$m ($H$ band) for the \amc~composition. 

\section{Scattering phase function of an optically thick disk: Effect of multiple scattering and limb brightening} \label{sec:appphase}

\begin{figure*}[t]
\begin{center}
\includegraphics[width=0.49\linewidth]{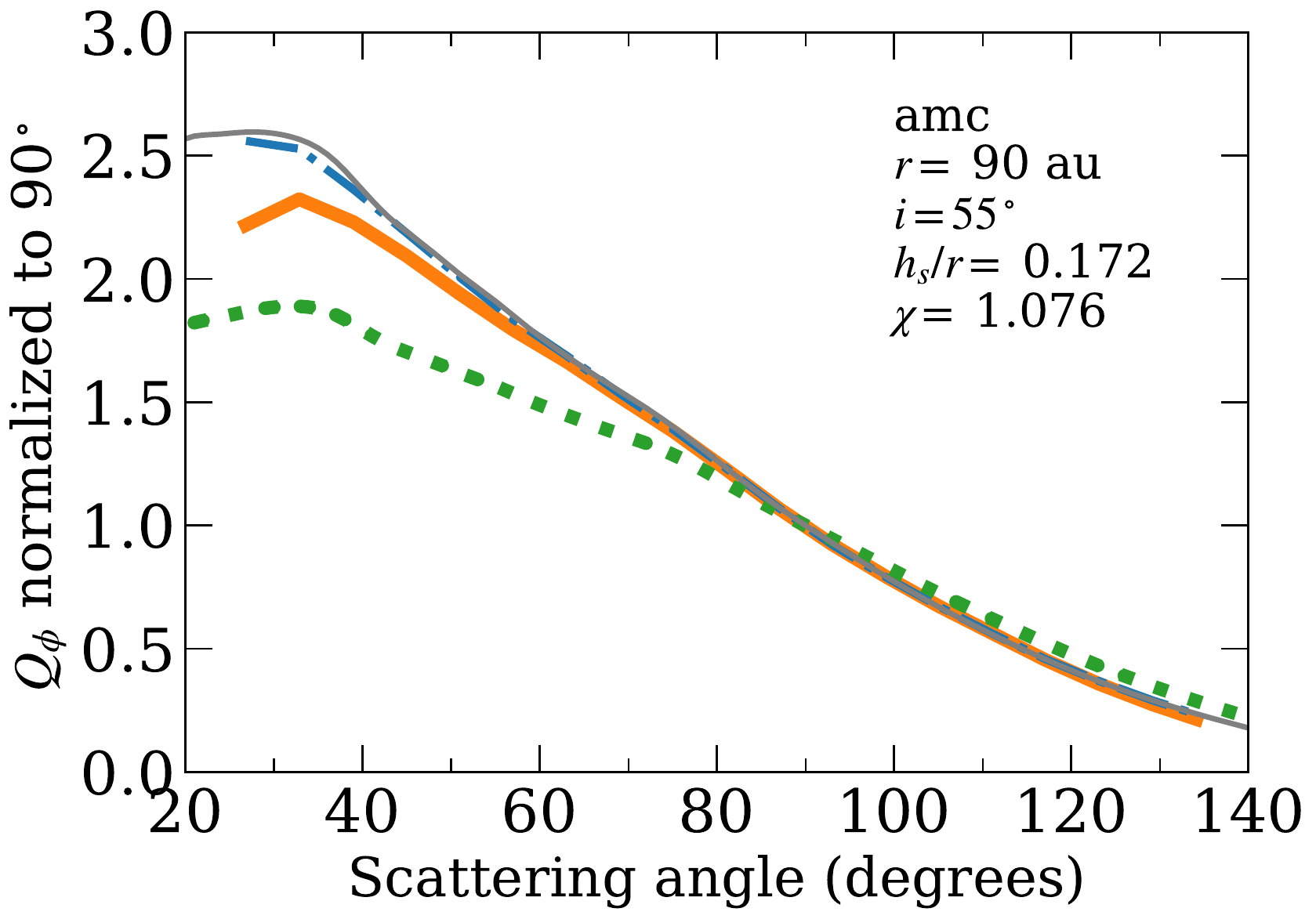} 
\includegraphics[width=0.49\linewidth]{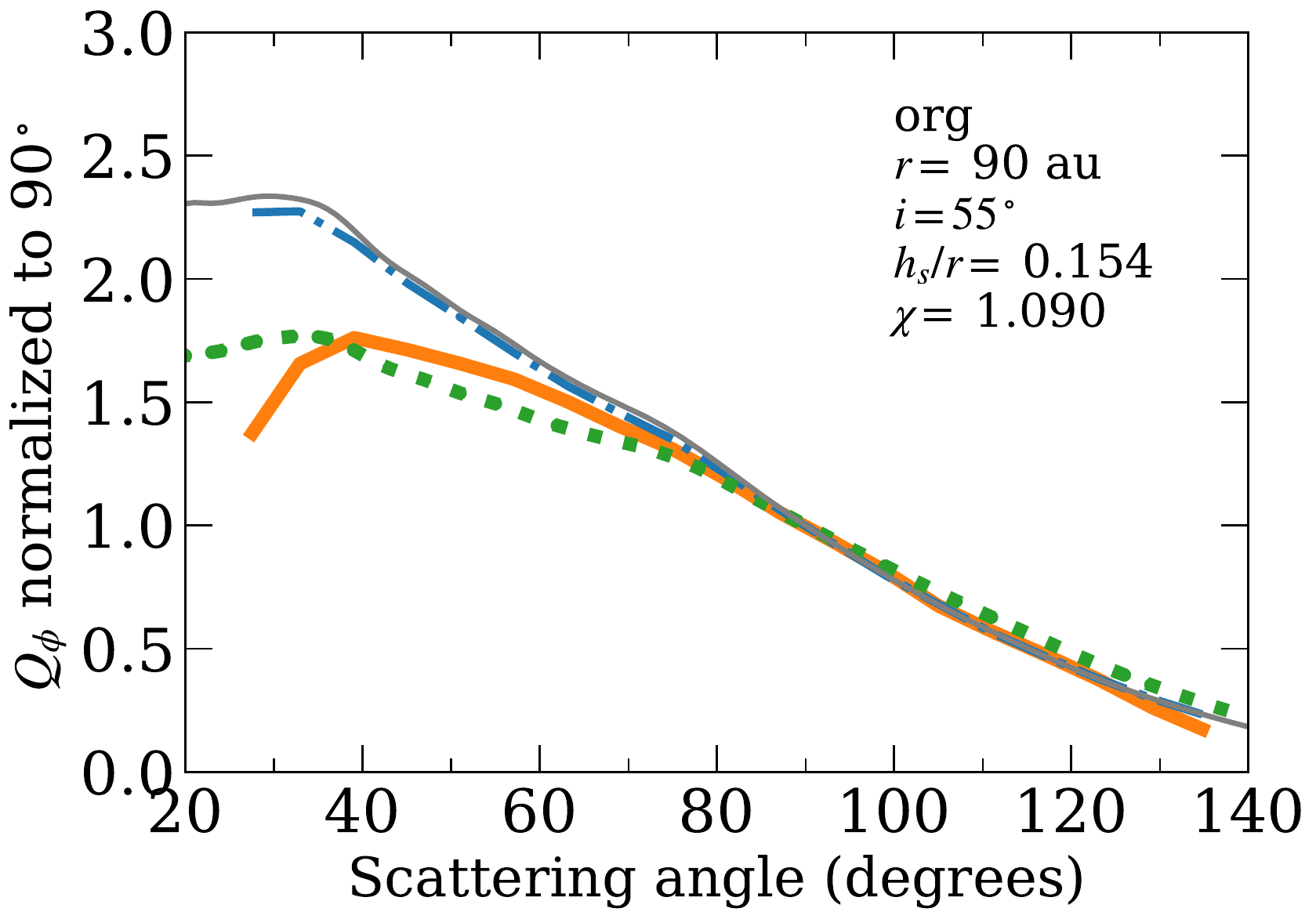} 
\includegraphics[width=0.49\linewidth]{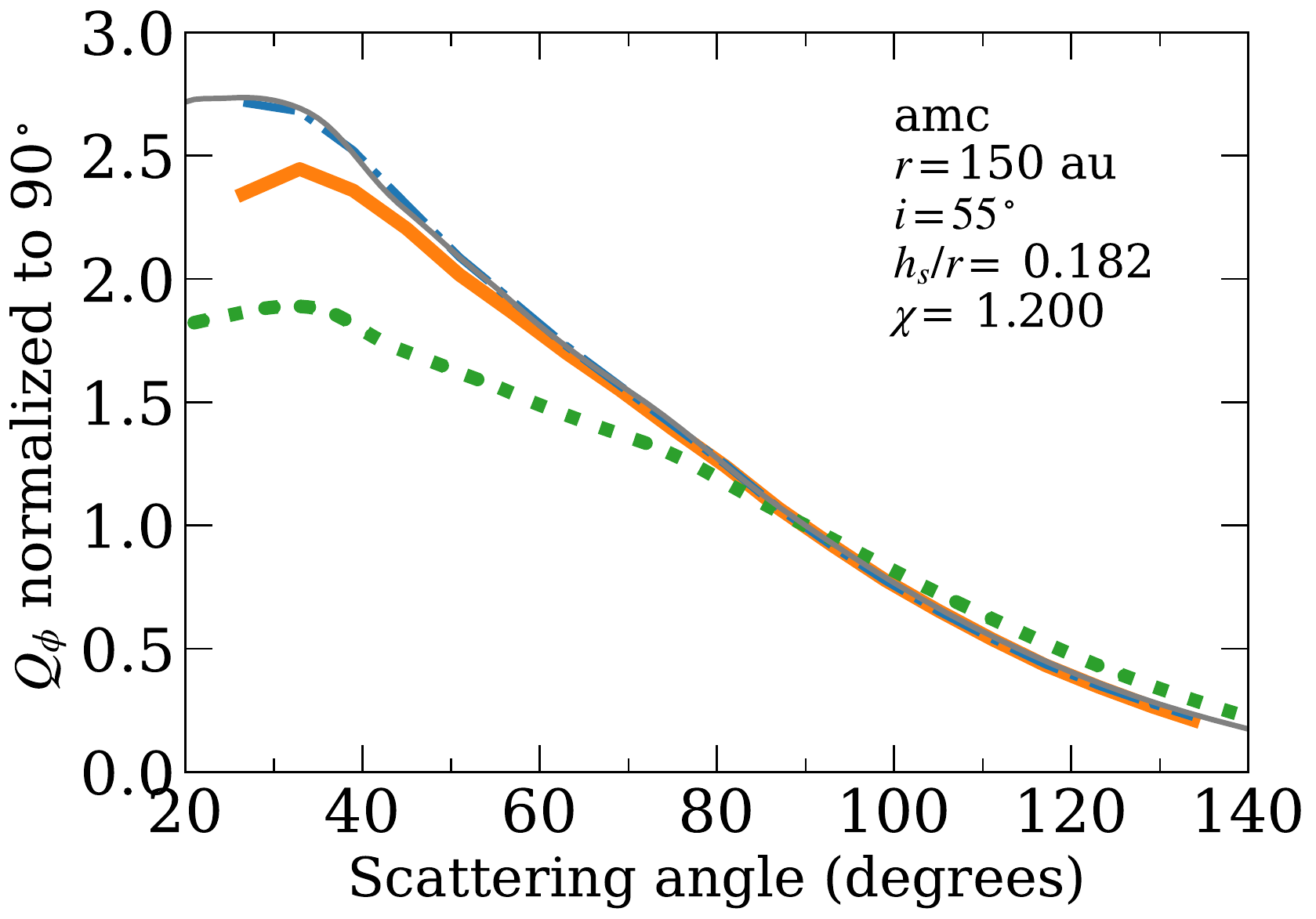} 
\includegraphics[width=0.49\linewidth]{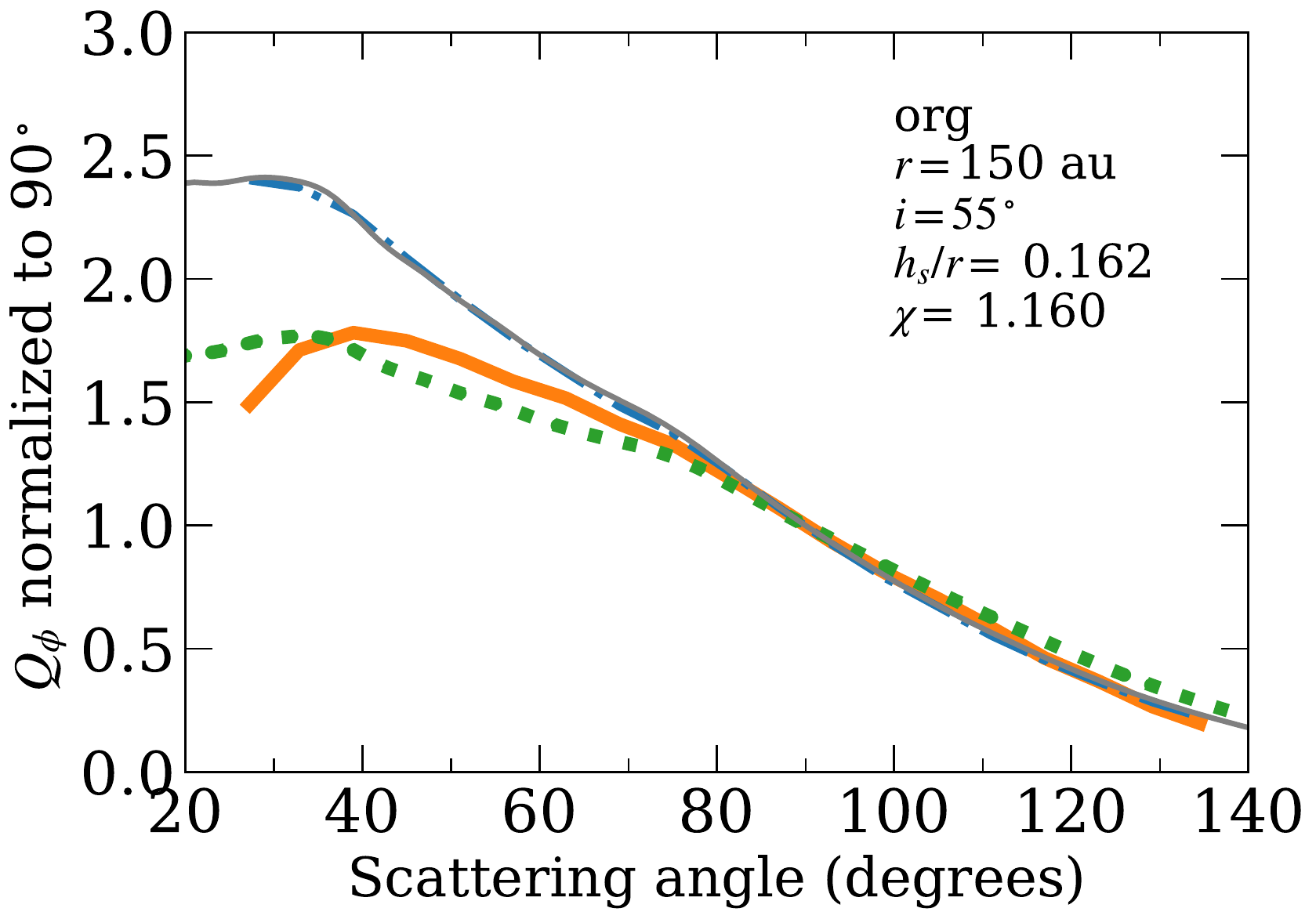} 
\caption{
Comparison of intrinsic and measured polarization phase functions, illustrating the effects of multiple scattering and limb-brightening, for two different disk locations (top row with $r=90$ au and bottom row with $r=150$ au) and two different material compositions (left column with highly absorbing amorphous carbon and right column with less absorbing organic carbon). In each panel, the thick dotted green line shows the intrinsic polarization phase function of \fan~with $N_\mathrm{max}=512$ with \amc200 (left panels) and \org200 (right panels) used in the radiative transfer model. That radiative transfer model was then used to produce an image. The solid orange line shows the apparent phase function measured from that image, which differs from the intrinsic function due to multiple scattering and limb brightening.  The dotted-dashed lines represent the phase functions obtained by measuring the images from a radiative transfer computation with multiple scattering switched off. The thin solid lines show the analytical estimate for $Q_\phi$ computed from the intrinsic phase function, ignoring multiple scattering and correcting for limb brightening using Equations (\ref{eq:qphi} and \ref{eq:mufactor}).
}
\label{fig:pfdisk}
\end{center}
\end{figure*}

\begin{figure}[t]
\begin{center}
\includegraphics[width=0.62\linewidth]{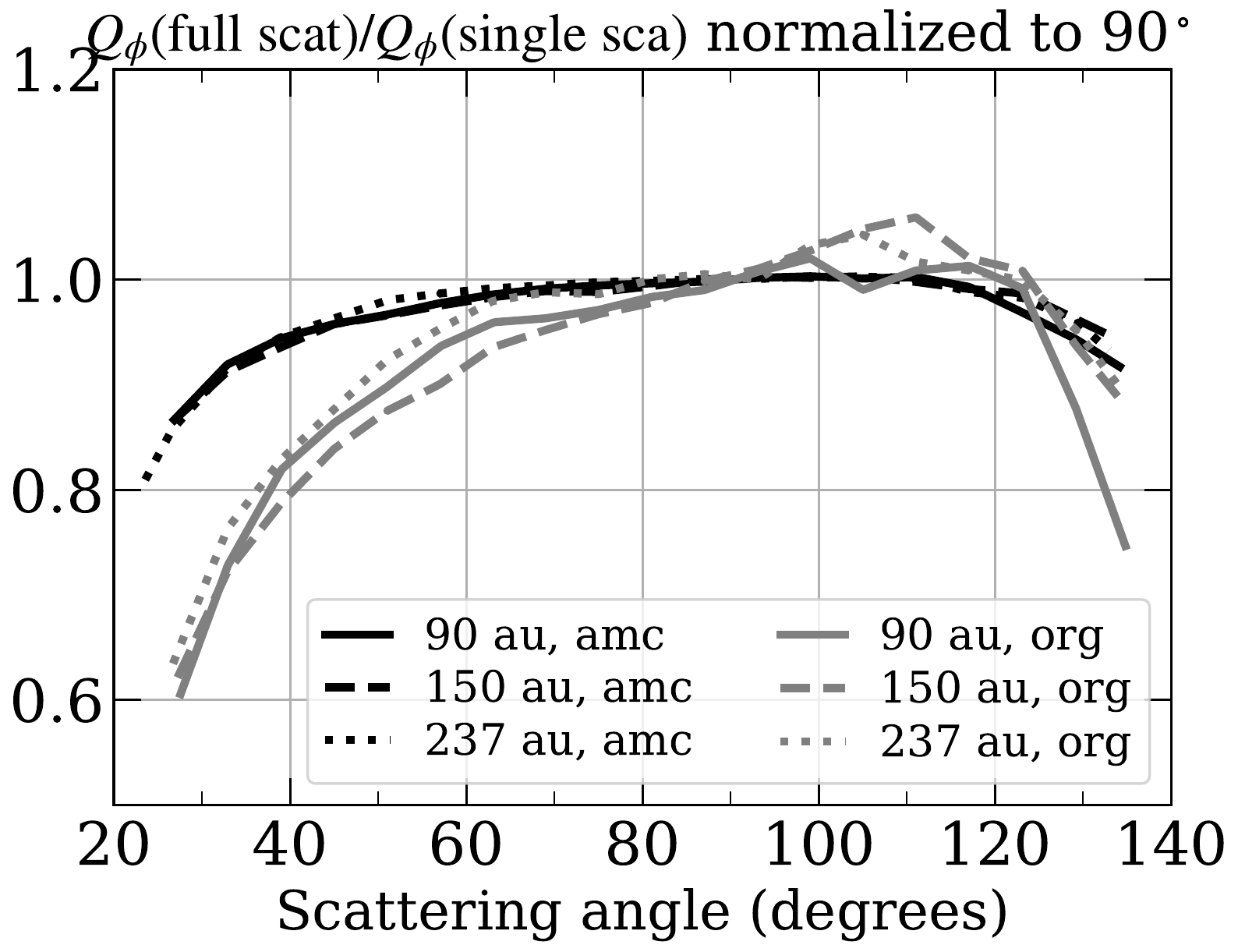}
\caption{The ratio of the polarization phase function extracted from model images with and without multiple scattering. The curves are normalized to a scattering angle of 90 degrees. The black and gray lines represent the results for the \amc~and \org~compositions, respectively. The solid, dashed, and dotted lines represent the results at 90 au, 150 au, and 237 au, respectively.}
\label{fig:multiplescat}
\end{center}
\end{figure}

\begin{figure}[t]
\begin{center}
\includegraphics[width=0.6\linewidth]{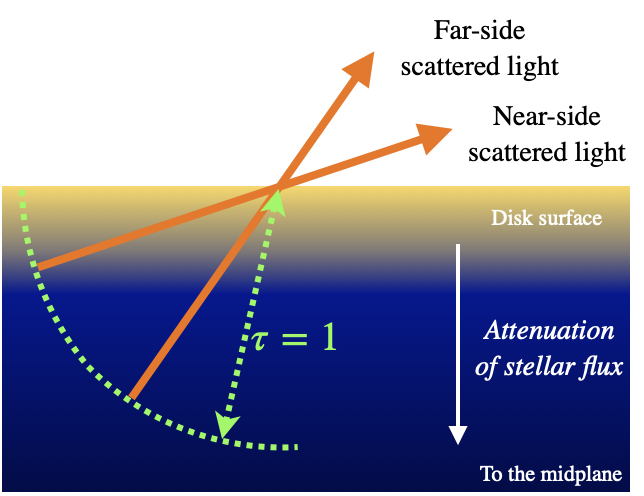}
\caption{A cartoon of a small patch of the disk surface illustrating the effect of limb brightening on the emergent intensity. The two solid arrows escaping from the disk surface represent scattered light for the near and far sides of the disk. The background color gradient represents the attenuation of stellar radiation from the top of the surface toward the midplane. The green dashed lines represent the location where the optical depth measured from the disk surface is unity. The line of sight for the near-side scattered light is exposed to a high stellar flux for a longer distance than that of the far side, and therefore the emergent intensity is enhanced.}
\label{fig:rtlim}
\end{center}
\end{figure}

\begin{figure*}[t]
\begin{center}
\includegraphics[width=0.4\linewidth]{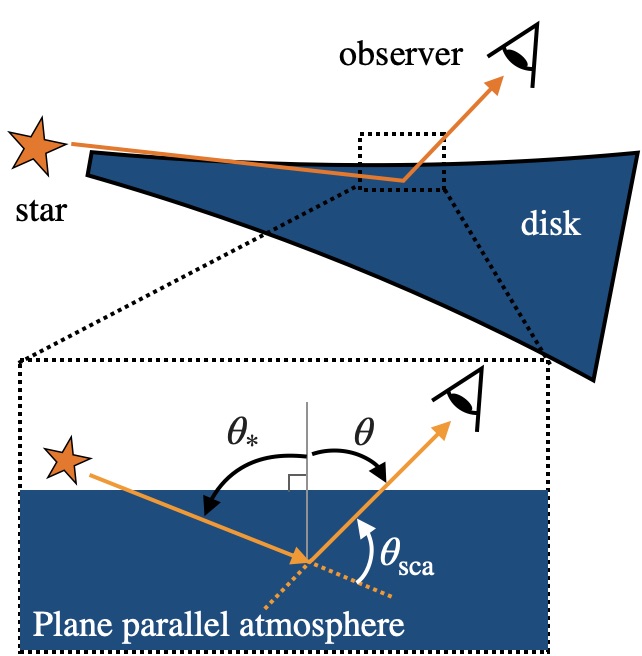}
\includegraphics[width=0.49\linewidth]{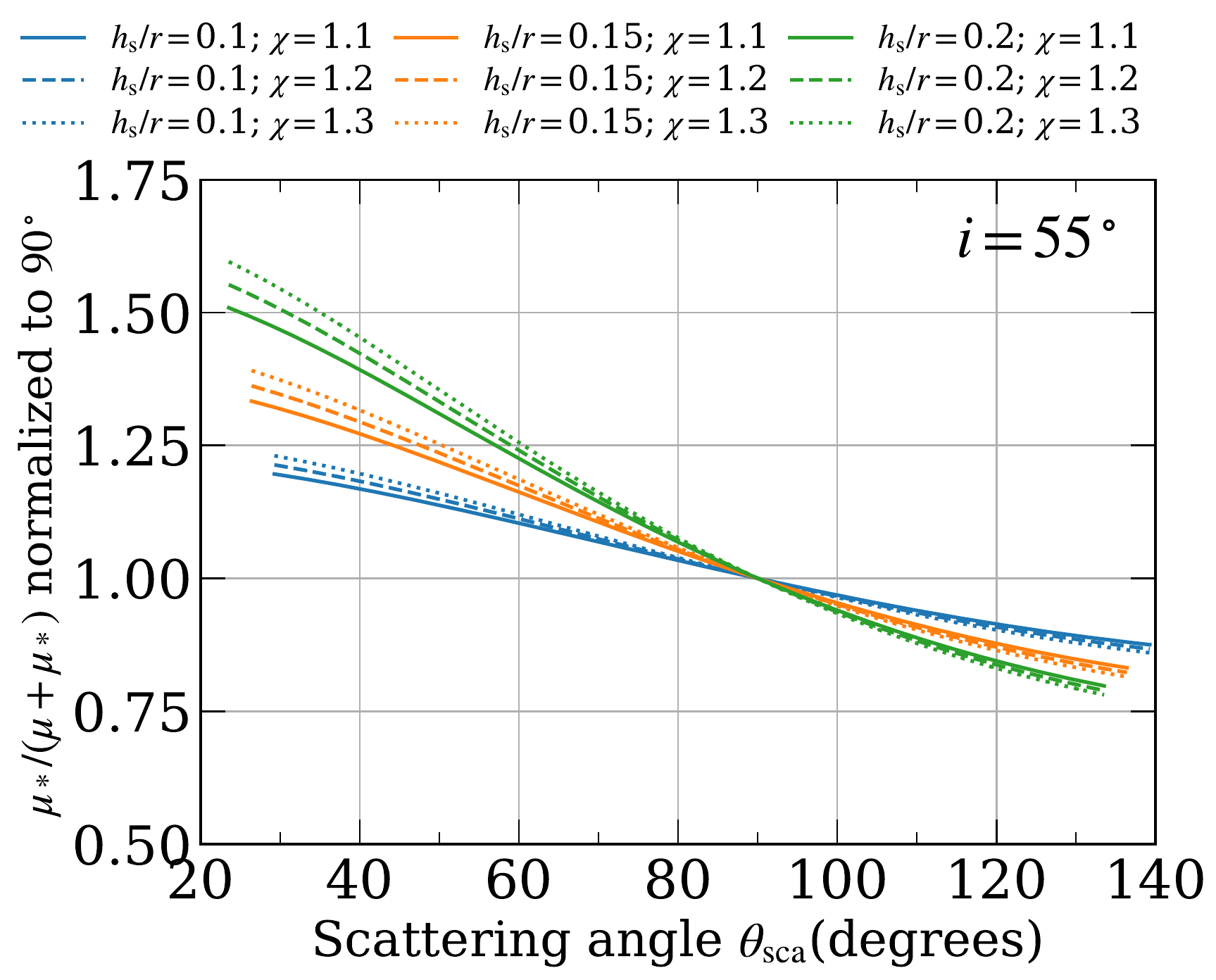}
\caption{A simple model to account for the effect of limb brightening on the phase function. The left cartoon summarizes the geometry of the plane-parallel atmosphere. The right panel shows the limb-brightening factor $\mu_*/(\mu_*+\mu)$ for various disk aspect ratios ($h_\mathrm{s}/r$) and flaring indices ($\chi$) at a disk inclination angle of $55^\circ$. The limb-brightening effect amplifies the near- and far-side brightness asymmetry of disk-scattered light.}
\label{fig:geom}
\end{center}
\end{figure*}

It is important to know if the phase function extracted from a scattered-light image of the disk is compatible with the scattering matrix element of each dust particle (the intrinsic phase function). For an optically thin and nonflat debris disk, \citet{Olofsson20} pointed out that the extracted phase function deviates from the intrinsic one because of the difference in the column density at each scattering angle. Here we study how the extracted phase function deviates from the intrinsic phase function for optically thick disks.

\subsection{Multiple scattering} \label{sec:multisca}
Figure \ref{fig:pfdisk} compares the extracted polarized phase functions and their intrinsic phase functions (correspond to $-S_{12}$ in \citet{Bohren83}). The assumed dust model is \fan~with $N_\mathrm{max}=512$ and either \org~or \amc. These plots demonstrate that the extracted phase functions do not necessarily coincide with the intrinsic phase function. 
One of the reasons is the occurrence of multiple scattering.
To study the impact of multiple scattering, we performed radiative transfer simulations by turning off multiple scattering and then extracted the phase functions (dotted-dashed lines in Figure \ref{fig:pfdisk}). Figure \ref{fig:multiplescat} shows the ratio of the polarization phase function with and without multiple scattering.
We found that multiple scattering affects the curves, particularly when each dust particle is made of weakly absorbing materials (the \org~composition). 
The effect of multiple scattering mainly appears at a smaller scattering angle and does not vary much among different disk radii (see Figure \ref{fig:multiplescat}).
Although multiple scattering plays a role to some extent, the extracted phase functions without multiple scattering still deviate from the intrinsic phase functions. Therefore, multiple scattering is not the only mechanism that causes a deviation from the intrinsic phase function. 

\subsection{Limb brightening} \label{sec:limb}

The reason why the polarization phase function still deviates from the intrinsic one, even in the absence of multiple scattering, is that the observer will see a slightly different disk height at each scattering angle (Figure \ref{fig:rtlim}). 
What we can observe is scattered light integrated along the line of sight of the observer. The line of sight for the near-side scattered light is exposed to a high stellar flux for a longer distance than that for the far side.
Because of the greater contribution of the stellar radiation to the emergent intensity, the near side becomes brighter than the far side {\it even if each dust particle obeys isotropic scattering}. We will refer to this effect as {\it the limb brightening}, as this effect may be regarded as the opposite effect to the limb darkening seen in the Sun.

To better illustrate the limb-brightening effect on the phase function, we consider a small patch of the disk surface so that we can apply a formula of diffuse reflection by a semi-infinite plane-parallel atmosphere (see Figure \ref{fig:geom} left).
Let us suppose a situation where the unpolarized stellar radiation is incident on a plane-parallel atmosphere, making an angle $\theta_*$ with respect to the surface normal ($0\le\theta_*<\pi/2$). The incident light will be scattered by dust particles in the atmosphere and then escape from it with an angle $\theta$ to the surface normal ($0\le\theta<\pi/2$). 
If the incident light has been scattered only once, the emergent total and polarized intensity can be expressed as \citep[e.g., Section 47.1 in][]{Changra1960}
\begin{eqnarray}
I(\mu,\mu_*)&=&\frac{F}{4\pi}\frac{\mu_*}{\mu+\mu_*}\omega_0P_\mathrm{11}(\mu,\mu_*),\\
Q_\phi(\mu,\mu_*)&=&\frac{F}{4\pi}\frac{\mu_*}{\mu+\mu_*}\omega_0P_\mathrm{12}(\mu,\mu_*), \label{eq:qphi}
\end{eqnarray}
where $F$ is the incident flux density, $\mu=\cos\theta$, $\mu_*=\cos\theta_*$, $\omega_0$ is the single scattering albedo of a dust particle, and $P_{11}$ and $P_{12}$ are the scattering phase functions for total and polarized intensity, respectively, and they are given by
\begin{eqnarray}
P_\mathrm{11}&=&\frac{4\pi}{\kappa_\mathrm{sca}}\frac{S_\mathrm{11}}{k^2 m},\\
P_\mathrm{12}&=&-\frac{4\pi}{\kappa_\mathrm{sca}}\frac{S_\mathrm{12}}{k^2 m},
\end{eqnarray}
where $\kappa_\mathrm{sca}$ is the scattering opacity, $k$ is the wavenumber, $m$ is the mass of a dust particle, and $S_{ij}$ is scattering matrix elements following the definition of \citet{Bohren83}. The total-intensity phase function ($P_{11}$) is normalized such that $\frac{1}{4\pi}\oint P_{11}d\Omega=1$. 
To calculate $\mu$ and $\mu_*$, we need to specify the disk flaring geometry. 
Suppose that the scattering surface obeys $h_\mathrm{s}\propto r^\chi$, we obtain

\begin{eqnarray}
\mu_*&=&\sin(\gamma'-\gamma),\label{eq:mustar} \\
\mu&=&\cos{i}\cos\gamma'-\cos\phi\sin\gamma'\sin{i}, \label{eq:mu}
\end{eqnarray}
where $\gamma=h_\mathrm{s}/r$, $\gamma'=dh_\mathrm{s}/dr=\chi h_\mathrm{s}/r$, $i$ is the disk inclination angle, and $\phi$ is the disk azimuthal angle; $\phi=0$ and $180^\circ$ correspond to the near and far sides of the disk, respectively. Similarly, we can express the scattering angle $\theta_\mathrm{sca}$ as \citep[e.g.,][]{Stolker16}:
\begin{equation}
\cos\theta_\mathrm{sca}=\sin{i}\cos\gamma\cos\phi+\cos{i}\sin\gamma. \label{eq:muscat}
\end{equation}
The angle range we can probe for the disk surface is $|i-\gamma'|\le \theta\le i+\gamma'$ and $\pi/2-(i+\gamma)\le\theta_\mathrm{sca}\le\pi/2+i-\gamma$.
Using Equations (\ref{eq:mustar}, \ref{eq:mu}, \ref{eq:muscat}) and assuming $i\ne0$, we obtain
\begin{equation}
\frac{\mu_*}{\mu_*+\mu}=\frac{\cos\gamma\sin(\gamma'-\gamma)}{\cos\gamma\sin(\gamma'-\gamma)+\cos{i}\cos(\gamma'-\gamma)-\sin\gamma'\cos\theta_\mathrm{sca}}, \label{eq:mufactor}
\end{equation}
Equation (\ref{eq:mufactor}) is a mildly decreasing function with scattering angle (Figure \ref{fig:geom} right). The larger the disk aspect ratio becomes, the steeper this factor becomes, although the effect of the flaring index $\chi$ is minor. 
Thus the disk aspect ratio is a key factor that makes the observed phase function steeper than the intrinsic phase function. Figure \ref{fig:pfdisk} shows 
the analytical estimate for $Q_\phi$ computed from the intrinsic phase function, ignoring multiple scattering and correcting for limb brightening
using Equations (\ref{eq:qphi} and \ref{eq:mufactor}). The disk geometry parameters ($\gamma$, $\gamma'$) are determined from the $\tau=1$ surface calculated for each model. It turns out that the simple model well reproduces the steep phase functions observed in the simulations without multiple scattering.

\subsection{The origin of 90 au and 150 au polarization phase functions} \label{sec:sim}
Figure \ref{fig:obs} shows that the 90 au and 150 au phase functions are similar to each other. Combining the results presented in Appendices \ref{sec:multisca} and \ref{sec:limb}, we conclude that the similarity of the polarization phase functions is likely attributable to the similar dust properties because (i) the degree of multiple scattering does not vary at different disk radii and (ii) these two surfaces have the similar aspect ratio ($h_\mathrm{s}/r\sim0.18$).


\end{document}